\documentclass[12pt,preprint2]{aastex}
\newcommand{\bee}{\begin{eqnarray}}
\newcommand{\ene}{\end{eqnarray}}
\newcommand{\half}{{1\over2}}

\newcommand{\nab}{\mbox{\boldmath{$\nabla$}}}
\newcommand{\vxi}{\mbox{\boldmath{$\xi$}}}
\newcommand{\vv}{\mbox{\boldmath{$v$}}}
\newcommand{\vB}{\mbox{\boldmath{$B$}}}
\newcommand{\vF}{\mbox{\boldmath{$F$}}}
\newcommand{\ve}{\mbox{\boldmath{$e$}}}
\newcommand{\vx}{\mbox{\boldmath{$x$}}}
\newcommand{\di}{\mbox{\rm div}}
\newcommand{\Y}{Y_\ell^m}

\shorttitle{Helioseismic Probing of Solar Variability}
\shortauthors{Dziembowski and Goode}

\begin{document}


\title{ Helioseismic Probing of Solar Variability:  The
Formalism and Simple Assessments}


\author{W. A. Dziembowski
} \affil{Warsaw University Observatory and Copernicus Astronomical
Center,Poland}

\and

\author{P. R. Goode
} \affil{Big Bear Solar Observatory, New Jersey Institute of
Technology,\\ Big Bear City, 92314, USA}
\email{pgoode@bbso.njit.edu}




\begin{abstract}

We derive formulae connecting the frequency variations in the
spectrum of solar oscillations to the dynamical quantities that
are expected to change over the solar activity cycle. This is done
for both centroids and the asymmetric part of the fine structure
(so-called even-$a$ coefficients).  We consider the near-surface,
small-scale magnetic and turbulent velocity fields, as well as
horizontal magnetic fields buried near the base of the convective
zone. For the centroids we also discuss the effect of temperature
variation.

We demonstrate that there is a full, one-to-one correspondence
between the expansion coefficients of the fine structure and those
of both the averaged small-scale velocity and magnetic fields.
Measured changes in the centroid frequencies and the even-$a$'s
over the rising phase solar cycle may be accounted for by a
decrease in the turbulent velocity of order 1\%. We show that the
mean temperature decrease associated with the net decrease in the
efficiency of convective transport may also significantly
contribute to the increase of the centroid frequencies.
Alternatively, the increase may be accounted for by an increase of
the small-scale magnetic field of order 100 G, if the growing
field is predominantly radial.

We also show that global seismology can be used to detect a field
at the level of a few times $10^5$ G, if such a field were present
and confined to a thin layer near the base of the convective
envelope.
\end{abstract}


\keywords{ Sun : Helioseismology, solar variability}


\section{Introduction}

We study global changes, over the solar cycle, in the sun's
eigenmode frequencies -- centroids and asymmetric fine structure
-- in a search of physical changes occurring beneath the
photosphere. There is abundant phenomenological information about
the helioseismic changes, but there is no satisfactory physical
model describing the changes. We consider three possible dynamical
sources of the evolution -- changes in the sub-photospheric
small-scale magnetic and velocity fields and a large-scale
toroidal field buried in a thin layer near the base of the
convection zone. Here, we develop the formalism needed to connect
these dynamical changes to frequencies changes.

In our treatment of the small-scale magnetic field, we generalize
the method of Goldreich et al.(1991, GMWK) to include the {\it
generalized} effect of the small-scale magnetic field on {\it
nonradial} modes, while further generalizing to a non-spherical
distribution of the averaged field. Although it is true that
radial modes may adequately represent lower degree (up to about
$\ell=60)$ nonradial p-modes, if the magnetic field effects were
confined to the outermost layers, this is not true for higher
degree p-modes or most f-modes. Still, the more important
generalization is that we treat a non-isotropic, non-spherical
field distribution, which allows us to interpret the
observed evolution of the anti-symmetric part of the fine
structure in the spectrum of solar oscillations (so-called
even-$a$ coefficients).

Furthermore, we study effects a small-scale, random velocity
field.  A role for the changing turbulent velocities has been
suggested by Kuhn (1999).  However, a first-principles treatment
still needs to be made. We give an estimate of the associated
temperature change and its effect on oscillation frequencies.

Finally, we con\-sider the ef\-fect of a buried toroidal field,
which may be expected to be confined near the base of the
convective envelope. The present work represents an advance over
earlier ones (Gough \& Thompson ,1990; Dziembowski \& Goode, 1991)
because we make a more explicit and useful formulation by
eliminating derivatives of the unknown dynamical quantities.  This
improved development allows us to obtain more physically revealing
formulae.  The application of this part of our work is determining
a stringent limit on the size of a buried toroidal field.

\section{The Helioseismic Data}

Solar frequency data are usually given in the form
\begin{equation}
\nu_{\ell n}^m=\bar\nu_{\ell n}+\sum_{k=1} a_{k,\ell n} {\cal
P}_k^{\ell}(m), \label{nu}
\end{equation}
where  the ${\cal P}$ are orthogonal polynomials (see Ritzwoller
\& Lavely 1991 and Schou et al. 1994). The remaining symbols ($n
\ell m$), in this equation have their usual meanings. This
representation ensures that $\bar\nu_{\ell,n}$ -- the centroid
frequencies -- are a probe of the spherical structure, while the
$a_{2k}$ -- the even-$a$ coefficients --  are a probe of the
symmetrical (about the equator) part of distortion described by
the corresponding $P_{2k}(\cos\theta)$ Legendre polynomials.  We
note that in lowest order, perturbations that are {\it
symmetrical} about the equator induce an {\it asymmetric} change
in the fine structure of the oscillation spectrum.

For the angular integrals, we have
\begin{equation}
Q^m_{k,\ell}\equiv\int_{0}^{2\pi}\int_{-1}^{1}|Y_{\ell}^{m}|^{2}
P_{2k}d\mu d\phi=S_{k,\ell}{\cal P}_{2k}^{\ell}(m) \label{Q}
\end{equation}
where $\mu=\cos\theta$ and
$$S_{k,\ell}=(-1)^k{(2k-1)!!\over k!}{(2\ell+1)!!\over(2\ell+2k+1)!!}
{(\ell-1)!\over(\ell-k)!}.$$

Following our earlier works (see e.g. Goode \& Dziembowski, 2002),
we use here the following convenient quantities, $\gamma_{k,\ell
n}$, through the following two relations,
\begin{equation}
\Delta\bar\nu_{\ell n}={\gamma_{0,\ell n}\over\tilde I_{\ell n}}
\label{D_nu_bar}
\end{equation}
and
\begin{equation}
a_{2k,\ell n}=S_{k,\ell}{\gamma_{k,\ell n}\over\tilde I_{\ell n}},
\label{a}
\end{equation}
where $\tilde I_{\ell,n}$ is the dimensionless mode inertia
calculated for our reference model.  A clear advantage of the
$\gamma$'s is that their growth replicates the growth of other
measures of solar activity. For the p-modes, the $1/\tilde I$
factor takes care of the $\ell$- and most of the $\nu$-dependence
in $\Delta\bar\nu$ and in the even-$a$ coefficients. The fact that
the residual $\nu$-dependence is weak points to a localization of
the source of the observed frequency changes close to to the
photosphere.

The numerical values of the $\gamma$'s scale with the square of
the eigenfunction normalization at the photosphere. The
normalization we adopted in our analyses of the SOHO MDI data
(e.g. Goode \& Dziembowski, 2002) and which is used throughout
present paper, is explained in the next section. With this
normalization, the value of $\gamma_0$ reaches up to the
$0.3\mu$Hz range. The absolute values of $\gamma_1$ and $\gamma_3$
are about twice larger. Having determined the set of $\gamma_k$,
one may construct seismic maps of the varying sun's activity
(Dziembowski \& Goode, 2002), that is the $\gamma(\mu)$
dependence. In such maps, a rising $\gamma$ reflects the local
rise of in the activity. The highest values of $\gamma(\mu)$ are
about $1 \mu$Hz and they are reached at $\mu\approx0.3$ and at the
peak of the activity. At activity minimum the highest
$\gamma(\mu)\approx0.2 \mu$Hz occurs in the polar region.

In the subsequent sections, we will connect the $\gamma$'s, to
magnetic and velocity fields that are expected to change in the
sun over its activity cycle.  To achieve this, we start from a
variational principle for oscillation frequencies. In our
expressions, the $\ell,n$ subscripts and the $m$ superscript will
not be given unless it is necessary for clarity.

\section{Variational principle for oscillation frequencies}

There are two ways of deriving the variational expression for
oscillation frequencies.  Both rely on the adiabatic
approximation, which is adopted throughout our study.  The first
of the two approaches begins with the linearized equations of
fluid motion about a steady configuration (see e.g Lynden-Bell \&
Ostriker, 1967, LBO).  The other uses Hamilton's principle (see
e.g. GMWK; Dewar, 1970). Here, we use the form given by LBO with
some simplification of the variational principle, while adding the
all-important contribution of the magnetic field, as calculated
explicitly by Dewar (1970).  The LBO form is valid for strictly
steady velocity fields. However, we make certain simplifications,
which will be explained later, to make it applicable to
statistically steady fields. With this, we write

\begin{equation}
\omega^2I=-\omega C+D \label{O}
\end{equation}
where
\begin{equation}
I=\int d^3\vx\rho|\vxi|^2 \label{I}
\end{equation}

\begin{equation}
C=2{\rm i}\int d^3\vx\rho\vxi^*\cdot(\vv\cdot\nab)\vxi, \label{C}
\end{equation}
and where $\vv$ represents the velocity field. The
eigen\-vec\-tors, in a spherically-symmetric and time independent
model of the sun, are expressed in the following standard  form
\begin{equation}
\vxi =r[y(r)\ve_r +z(r)\nab_H]\Y(\theta,\phi) \exp(-{\rm i}\omega
t). \label{xi}
\end{equation}

We adopt some approximations regarding the eigenfunctions. In
addition to adiabaticity, we assume the Cowling approximation is
valid, which is well-justified in our application to solar
oscillations. Further, we will make use of the fact that the
oscillations are either of high degree or high order, which means
that
$$|\vxi|\ll{\rm Max}(r|\vxi_{;r}|,\ell|\vxi|).$$
Equivalent approximations were also made by GMWK but, in addition,
they ignored the angular dependence of the displacement.

Like  LBO, we separate the various contribution to $D$,
\begin{equation}
D=D_p+D_g+D_v+D_M. \label{D}
\end{equation}
The pressure term,
\begin{equation}
D_p=\int d^3\vx p[\Xi+(\Gamma-1)|\di\vxi|^2], \label{D_p}
\end{equation}
is the same as in LBO. The quantity $\Gamma$, usually denoted as
$\Gamma_1$, is the adiabatic exponent. The quantity $\Xi$ is a
completely contracted double dyadic product,

\noindent $\nab\vxi^*:\nab\vxi$. Adopting the standard summation
convention, we have
$$\Xi=\xi_{j;k}^*\xi_{k;j}=(\xi_j^*\xi_{k;j}-\xi_k^*\di\vxi)_{;k}+|\di\vxi|^2,$$
where the subscript ``;'' denotes covariant derivatives. However,
with our approximation regarding $\vxi$, contributions from the
terms involving the Christoffel symbols are negligible, and the
derivatives may be regarded as component derivatives. In terms of
the radial eigenfunctions, $y$ and $z$, with the adopted
approximations we have \bee
\Xi&\approx&[r(yry_{;r}-y\lambda)_{;r}+\lambda^2]|\Y|^2+
\nonumber\\&&\hspace{-0.8cm}r(yz)_{;r}|\nab_H\Y|^2+[0.5(yrz_{;r}-z\lambda)
|\Y|^2_{;\theta}\nonumber\\&&\hspace{-0.8cm}
+z^2\nab_HY^{m*}_\ell\cdot\nab_HY^m_{\ell;\theta}]_{;\theta}+[...]_{;\phi},
\nonumber\ene
 where $$\lambda=y{gr\over c^2}-z{\omega^2r^2\over
c^2}$$ is radial eigenfunction corresponding to $\di\vxi$, $g$ is
the local gravity, and $c$ is the speed of sound. The last term in
$\Xi$ is obtained from the preceding one by the replacement $\theta\leftrightarrow\phi$.
Further, in the adopted approximation, we have
\begin{equation}
ry_{;r}=\lambda +\Lambda z, \label{ry}
\end{equation}
where $\Lambda=\ell(\ell+1)$ and
\begin{equation}
rz_{;r}=y-\lambda\left({N\over\omega}\right)^2{c^2\over gr}.
\label{rz}
\end{equation}
The term containing the Brunt-V\"ais\"al\"a frequency, $N$, is of
the same order as the first one for p-modes only, and only in the
outermost layers. However, the whole contribution from the term
involving $rz_{;r}$ is small. Hence, we will ignore the term, so
that

\bee \Xi&=&\lambda^2|\Y|^2+(y^2+\lambda z+\Lambda z^2)
\nonumber\\&&\hspace{-0.8cm}(\Lambda|\Y|^2+|\nab_H\Y|^2)+
\nonumber\\&&\hspace{-0.8cm}
[...]_{;\theta}+[...]_{;\phi}. \label{Xi} \ene
The explicit
expressions for the last two terms will not be needed.

The gravity term simplifies to
\begin{equation}
D_g=-2\int d^3\vx\rho{g\over r}|\xi_r|^2 \label{D_g},
\end{equation}
after using the Cowling approximation, while the velocity term,
\begin{equation}
D_v=-\int d^3\vx\rho|(\vv\cdot\nab)\vxi|^2, \label{D_v}
\end{equation}
is the same as in LBO, where it was derived for a steady field
velocity field. We will use the same form in our application to a
statistically steady turbulent field. The expression for the
magnetic term, which is taken from Dewar (1970), is

\bee D_M&=&{1\over4\pi}\int d^3\vx\bigg[|(\vB\cdot\nab)\vxi|^2-
\nonumber\\&&2\di\vxi^*\vB\cdot(\vB\cdot\nab)\vxi+
\nonumber\\&&\half|\vB|^2(\Xi+|\di\vxi|^2)\bigg] \label{D_M} \ene

We now perturb eq.\,[\ref{O}] about the static, non-magnetic
equilibrium state. The $\Delta$ denotes changes in parameters
relative to this state.  However, for centroid frequencies,
$\Delta$ is defined with respect to activity minimum because we do
not have models of the sun predicting centroid frequencies with
$\mu$Hz precision.  Since we want to consider terms that are
quadratic in velocity, in principle, we need to consider a second
order perturbational expression, which is
\bee \Delta\omega&=&-{C\over2I}+{C^2\over8I^2\omega}-{\omega\Delta C\over2I\omega}
\nonumber\\&&+{\Delta D_s+D_v+D_M\over2I\omega}\ene
where
\begin{equation}
\Delta D_s=\Delta(D_p+D_g)-\omega^2\Delta I. \label{Delta_D_s}
\end{equation}
Actually, we do not calculate the $C$ integral or its
perturbation, but only comment on the role of the terms in $C$ for
various velocity fields. The first term, which is linear in
velocity, results from rotation, and gives rise to odd-$a$
coefficients, which we are not treating here.  It may be easily
shown that both meridional and statistically steady turbulence do
not contribute. The second term, $\propto C^2$ due to rotation
gives a negligible contribution ( Dziembowski and Goode, 1992) to
p-mode splitting. The term $\propto\Delta C$ arises from the first
order perturbation of the eigenfunctions due to the velocity
fields.  Here the contribution from rotation and meridional
circulation can be shown to be negligible. The only quadratic
effect of rotation, which we found to be significant for p-modes
is that of the centrifugal distortion.  Thus, it is  included in
the $D_s$ term. For the f-mode even-$a$'s, which are not
accurately determined, the terms involving $C^2$ may be important.
The alternative approach, which has been used by us in all our
analyses of the even-$a$ coefficients, is to evaluate the
centrifugal contribution and subtract it from the data. We neglect
contribution from turbulence to the term $\propto\Delta C$,
because we include only effects of interaction of oscillations
with the averaged velocity fields. It has to be kept in mind,
however, that not all effects of turbulence are included in our
formalism. So that we are left with the expression
\begin{equation}
\Delta\omega={1\over2I\omega}\left(\Delta D_s+D_v+D_M\right)
\label{D_o1}
\end{equation}
In the  $D_v$ term, we consider only effects of the turbulent
pressure and we will be interested in the part that may vary with
the solar activity. We do not have yet observational evidence for
changes in turbulent velocity but such changes are expected. The
only global changes in velocity which were definitely detected are
the torsional oscillations but their effect on frequencies we
estimated to be insignificant.

The $D_s$ integral may be calculated considering either Eulerian
or Lagrangian perturbations. The results must be the same. The
$D_v$ and $D_M$ integrals are treated as perturbations.  We may
see that the integrands do not involve differentiation of the
unknown characteristics of the velocity and magnetic fields,
rather the differentiation is placed upon the eigenfunctions,
which are known. This is clearly advantageous and we will apply
the same strategy in the evaluations of $\Delta D_s$.

The calculated frequency perturbation for individual ($\ell n
m$)-modes are linked to the $\gamma's$ defined in  eqs.\,[\ref{D_nu_bar}] and
[\ref{a}]
by the following relation
\begin{equation}
\Delta\omega_{\ell n}^m={2\pi\over\tilde I_{\ell n}}\sum_{k=0}\gamma_{k,\ell
n}Q^m_{k,\ell} ,
\label{D_o2}
\end{equation}
with
$$I_{\ell n}=R_\odot^5\bar\rho_\odot\tilde I_{\ell n},$$
$$\tilde I_{\ell n}=\int dx x^4\tilde\rho{\cal E},$$
$$ x={r\over R_\odot},\qquad \tilde\rho={\rho\over\bar\rho},$$
and
$${\cal E}=y^2+\Lambda z^2.$$
As a normalization of the eigenfunctions, we adopt
$$y_{\ell,n}(r_{\rm phot})=2\times10^4.$$
This is an arbitrary choice which leads to maximal $\gamma$'s in
the $(0.2 - 1)\mu$Hz range, which is the same order as the
frequency shifts. This normalization is assumed in all expressions
for $\gamma$ provided in this paper.

\section{Dynamical perturbations of the structure}

Here, we include the dynamical effects of the magnetic $\vB$, and
those of the velocity fields, $\vv$. We write the condition of
mechanical equilibrium, in the presence of perturbing force $\vF$,
in the following form,

\begin{equation}
\nabla p+\rho g\ve_r=\vF, \label{nabla_p}
\end{equation}
where
\bee \vF &\equiv&-\left({\partial{\cal V}\over\partial
r}\ve_r+ {\partial{\cal H}\over\partial\theta}{\ve_\theta\over
r}\right)=\\
&&\hspace{-1cm}-{1\over4\pi}\left[\half\nabla\vB^2
-(\vB\cdot\nabla)\vB\right]-\rho(\vv\cdot\nabla)\vv \nonumber
\label{F} \ene

We neglected the perturbation of the gravitational
potential, which is justified
 because we considering perturbing forces concentrated
in thin layers containing little mass. eq.\,[\ref{nabla_p}]
implies
\begin{equation}
p=-{\cal H}+h(r) \label{p}
\end{equation}
and
\begin{equation}
\rho={1\over g}{\partial\over\partial r}({\cal H}-{\cal V}-h).
\label{rho}
\end{equation}
The quantities ${\cal V}$ and ${\cal H}$ represent non-gas
pressures, which in general are anisotropic. It should be noted
that when the non-gas pressure is iso\-tro\-pic, then the mass
distribution remains spherically symmetric. The quantity $h(r)$
may only be determined by utilizing the condition of thermal
equilibrium.  For the non-spherically symmetric parts of the
force, the pressure and density follow from the condition of
mechanical equilibrium.

From now on, we treat $\vF$ as a small perturbing force.
Primed letters denote Eulerian perturbations of the respective structure
parameters, letters preceded by $\delta$ denote Lagrangian
perturbations, and letters without such symbols imply unperturbed
variables. We use the standard relation $$\delta f= f'+{df\over
dr}\delta r$$ and  we adopt $\delta M_r=0$, for both spherical and
aspherical perturbations.

Note that if we make a Legendre expansion in even orders,
$P_{2k}(\cos\theta)$, of ${\cal H}$ and ${\cal V}$, all the
expansion coefficients ${p'_k}$ and ${\rho'_k}$, starting from
$k=1$ are completely specified. Here we are considering only even
order polynomials because these are the ones that contribute to
the even-$a$ coefficients. The anti-symmetric (odd-order)
polynomials average out. From eqs.\,[\ref{p}] and [\ref{rho}], we
get for $k>0$,
$$p'_k=-{\cal H}_k$$
and
\begin{equation}
\rho'_k={1\over g} {d\over dr}({\cal H}_k-{\cal V}_k). \label{p_k}
\end{equation}
The expansion coefficients for the Lagrangian perturbations are
calculated as follows. From the radial component of
eq.\,[\ref{nabla_p}] we have
\begin{equation}
\delta_k p=-{\cal V}_k+4\int_r^R{\delta_kr\over r}g\rho dr
\approx-{\cal V}_k\label{del_p_k}
\end{equation}
and from mass conservation
\begin{equation}
{d \delta_k r\over dr} =-{\delta_k\rho\over\rho}-2{\delta_kr\over r}
\approx-{\delta_k\rho\over\rho}.
\label{delta_0_r}
\end{equation}

The approximate equality in the preceding equation corresponds to
neglecting of the perturbation in the mass distribution above the
point under consideration. This is certainly valid for all of our
applications. The approximate equality in eq.\,[\ref{delta_0_r}]
is just the local plane parallel approximation. This is valid for
most of applications considered here. The only possible exception
will be discussed briefly in subsection 5.1. Both approximations
were adopted in GMWK. We stress, however, they are not needed for
deriving expressions for $\gamma_k$, except for $k=0$. For $k>0$,
it is only important and, in fact well justified, for seismic
determination of the aspherical part of the subphotospheric
temperature changes. To this aim, we first derive an expression
for $\delta_k r$ from the relation between $p'_k$ and $ \delta_k
p$,
\begin{equation}
\delta_k r={{\cal V}_k-{\cal H}_k\over g\rho}.
\end{equation}
Then, using the linearized $p(\rho,T)$ relation we obtain \bee
{\delta_k T\over T}&=&-{1\over\chi_Tp}\bigg[{\cal
H}_k+\Big(\chi_\rho{d\over d\ln p}+ \nonumber\\&& \chi_T{d\ln
T\over d\ln p}\Big)({\cal V}_k-{\cal H}_k)\bigg].
\label{delta_0_T}\ene Here, we used a standard notation in
astrophysics, e.g. $\chi$'s denote derivatives of $\log p$ with
respect of the $\log \rho$ and $\log T$.  We see that for the
non-spherical part, all perturbations of thermodynamical
quantities are determined by ${\cal H}$ and ${\cal V}$. This is
not true for $k=0$, where one of the thermodynamical parameters is
left free. Choosing the temperature, we have
\begin{equation}
{\delta_0\rho\over\rho}=-{1\over\chi_\rho}\left({{\cal
V}_0\over p}+ \chi_T{\delta_0 T\over T}\right)
\label{delta_0_rho_T},
\end{equation}
or, if we choose the entropy per mass variation, $\delta_0 S$, in place of
$\delta_0 T$,
\begin{equation}
{\delta_0\rho\over\rho}=-{1\over\Gamma}{{\cal V}_0\over
p}-{\chi_T\over\chi_\rho} {\delta_0 S\over c_p}.
\label{delta_0_rho_s}
\end{equation}

\subsection{Turbulent pressure}
The large-scale average of the Reynold's stress,
$\vF=-\overline{\rho(\vv\cdot\nab)\vv}$, due to the turbulent
velocity, is evaluated in the local Cartesian system with axes
parallel  to $\ve=(e_r,e_\theta,e_\phi)$ (effects of curvature are
negligible for this small-scale velocity field). Then, we have
$$F_i=-\overline{(\rho v_jv_i)}_{;j}+\overline{(\rho
v_j)_{;j}v_i}.$$ We use the following relations
$$\overline{\rho v_jv_i}=\delta_{ji}\overline{\rho v_i^2}$$
and
$$\overline{(\rho
v_j)_{;j}v_i}=-\overline{{\partial\rho\over\partial t}v_i}=0, $$
where double-subscripted $\delta$ is, as usual, the Kronecker
symbol. That is, we assume uncorrelated velocity components and
the rate of density fluctuations.
 Hence, we have
\begin{equation}
F_i=-\overline{(\rho v_i^2)}_{;i}\quad\mbox{ no summation over $i$
!} \label{F_i}
\end{equation}
Further, we allow the vertical component ($r$) to be statistically
different from the two horizontal ($\theta$ and $\phi$) ones. In
this, values of $\overline{\rho v_j^2}$ are treated as functions
of depth, and slowly varying functions of the co-latitude. The
latter dependence is represented in the form of a Legendre
polynomial series, \bee \overline{\rho
v_iv_j}&=&\rho\delta_{ij}\sum_{k=0} [\delta_{jr}{\cal
T}^V_k(r)+\nonumber\\&&\hskip -1.3cm\half{\cal
T}^H_k(r)(\delta_{j\theta}+ \delta_{j\phi})]P_{2k}(\cos\theta),
\label{rho_v_iv_j} \ene where we included only terms that are
symmetric about equator.

Inserting this expression into eq.\,[\ref{F_i}] and using the definition of
$\cal V$ and $\cal H$ as given  in eq.\,[21], we get
\begin{equation}
{\cal V}_k=\rho{\cal T}^V_k, \qquad{\rm and}\qquad {\cal
H}_k=\half\rho{\cal T}^H_k. \label{V_k_v}
\end{equation}

\subsection{Small-scale random magnetic field}

Our treatment of the small-scale magnetic field is analogous to
that of the turbulent velocities. That is, the correlation matrix
for the field components is represented in the form of the
following Legendre polynomial series,
\bee
\overline{B_iB_j}&=&\delta_{ij}\sum_{k=0} [\delta_{jr}{\cal
M}^V_k(r)+\nonumber\\&&\hskip -1.3cm\half{\cal M}^H_k(r)(\delta_{j\theta}+
\delta_{j\phi})]P_{2k}(\cos\theta). \label{B_iB_j}
\ene
Components of the Lorentz force, treated locally as Cartesian, are
given by
$$F_i={1\over4\pi}\sum_{j\ne i}\left(B_jB_{i;j}-\half (B^2_j)_{;i}\right).$$
Averaging over wide zonal areas and making use of
$B_{i;i}=0$, we get
$$\overline{B_\theta B_{r;\theta}+B_\phi B_{r;\phi}}=
\half{\partial\overline{B_r^2}\over\partial r}.$$
Thus,
$$F_r={1\over8\pi}{\partial\over\partial
r}\left(\overline{B_r^2}-\overline{B_H^2}\right),$$ which is the
same expression that was obtained by GMWK. Let us note that the
net effect of the vertical component of the random field magnetic
on the vertical structure is opposite to that of the horizontal
components. The radial component acts as a negative pressure, when
it rises the gas pressure must rise too.

To evaluate the horizontal
force, we use
$$\overline{B_\phi B_{\theta;\phi}}=\half(\overline{B_\theta^2})_{;\theta}+
\overline{B_{r;r}B_\theta},$$ to get
\bee F_\theta&=&{1\over4\pi}\bigg[\overline{(B_rB_\theta)}_{;r}+\half\nonumber\\&&
\bigg(\overline{B_\theta^2}-\overline{B_\phi^2}-\overline{B_r^2}\bigg)_{;\theta}\bigg]
=-{\overline{(B_r^2)}_{;\theta}\over8\pi}.\nonumber
\ene Finally, for the coefficients in the expansion of the
magnetic pressure, we obtain
\begin{equation}
{\cal V}_k={{\cal M}^H_k-{\cal M}^V_k\over8\pi} \quad {\rm
and}\quad {\cal H}_k={{\cal M}^V_k\over8\pi}.  \label{V_k_B}
\end{equation}
Just as in the case of the turbulent velocity field (see
eq.\,[\ref{V_k_v}]), isotropy implies ${\cal V}_k={\cal H}_k$,
hence no density perturbation, but effects of a departure from
isotropy are clearly different.
\subsection{Large-Scale Toroidal magnetic field}
The large-scale field, $\vB=B_{t}(r,\theta)\ve_\phi$, gives rise to the
Lorentz force
$$\vF=-{\nab (B^2_{t} r^2\sin^2\theta)\over8\pi r^2\sin^2\theta}
\approx-{\nab (B^2_{t} \sin^2\theta)\over8\pi\sin^2\theta}.$$
The last approximation is valid for a field confined to a narrow
layer, which we will assume here, so that we have
$${\cal V}={B^2_{t}\over 8\pi}$$
 and
$${\cal H}={1\over8\pi}\int{d\theta\over\sin^2\theta}
{\partial(B^2_{t}\sin^2\theta)\over\partial\theta}.$$

We now put $B_t(r,\theta)$ in the form of the following series,
\begin{equation}
B_t(r,\theta)=\sum_j\sqrt{2j+1\over2j(j+1)}B_{t,j}(r){dP_j(\cos\theta)\over
d\theta}. \label{B_t}
\end{equation}
Note that with this representation, $B_{t,j}$ is the surface
averaged intensity of the field component at a distance $r$ from
the center. Considering only first two terms in the expansion, we
get the following non-zero components of the Legendre polynomials
expansion for ${\cal V}$

\begin{equation}
{\cal V}_0={B^2_{t,1}+B^2_{t,2}\over 16\pi},\label{V_0_B_t}
\end{equation}
$${\cal V}_1={1\over 16\pi} \bigg(-B_{t,1}^2
+{5\over7}B^2_{t,2}\bigg),$$
\begin{equation}
{\cal V}_2=-{3B^2_{t,2}\over
28\pi}, \label{V_1_B_t}
\end{equation}
and for ${\cal H}$
$${\cal
H}_1=-{1\over8\pi}\left(B_{t,1}^2+{5\over7}B^2_{t,2}\right),$$
\begin{equation}
{\cal H}_2=-{9B^2_{t,2}\over 56\pi}. \label{H_1_B_t}
\end{equation}
\section{The term arising from the structural perturbation, $\Delta D_s$}
The frequency perturbation arising through the perturbation of the
structure for all forces considered by us is, typically, of the
same order as that arising directly from the forces. We will
express now perturbation the structural term $D_s$ in terms of
${\cal V}$ and ${\cal H}$ calculated in the previous section. The
variational principle ensures that we may keep $\vxi$ (not $y$ and
$z$ !) unperturbed.
\subsection{Calculations of $\Delta D_s$ for centroid frequency shift}

Here, using the Lagrangian formulation of the perturbations is
more convenient. Since we have $\delta(dr r^2\rho)=0$ (we drop
from here on the subscript at $\delta$ if it is zero), there is no
contribution from $\Delta I$. Furthermore, with our approximation
for the eigenfunctions, the contribution from $D_g$ is negligible.
For the present application, it is convenient to write
 eq.\,[\ref{D_p}] in the form,
$$ D_p=\int dr r^2p\left({d{\cal Z}\over dr}+\lambda^2\right),$$
where ${\cal Z}=2r(\Lambda yz-y^2)$ and, which after integration
by parts becomes $$ D_p=\int dr r^2\left[\left(g\rho-{2p\over
r}\right){\cal Z}+p\Gamma\lambda^2\right].$$ In the whole solar
envelope the second term in the coefficient at ${\cal Z}$ is much
less than the first one and it will be ignored. Now we calculate
$\Delta D_{s,0}\approx\Delta D_p$ using $${\delta g\over
g}=2{\delta{\cal Z}\over{\cal Z}}=-2{\delta r\over r}$$ and
$${\delta(c^2)\over c^2}={\delta p\over
p}(1+\Gamma_p)-{\delta\rho\over\rho}(1-\Gamma_\rho),$$ where we
denoted by $\Gamma_p$ and $\Gamma_\rho$ logarithmic derivatives of
$\Gamma$.
Further, we use eq.\,[\ref{del_p_k}] to eliminate $\delta p$ and eq.\,[\ref{delta_0_rho_T}]
to eliminate $\delta\rho$. Finally, with our approximations regarding the
eigenfunctions, we get
\bee
\Delta D_{s,0}&=&\int dr r^2\bigg[{\cal D}_{\rm isoth}{\cal V}_0+\nonumber\\&&
p\left({\cal D}_T{\delta T\over T}+{\cal D}_r{\delta r\over
r}\right)\bigg]\label{D_s_01} \ene
where
\bee
{\cal D}_{\rm isoth}&=&-\Gamma\bigg[\bigg(1+\Gamma_p+{1+\Gamma_\rho\over\chi_\rho}\bigg)\lambda^2
\nonumber\\&&+{2\Lambda z\lambda\over\chi_\rho}\bigg],\label{D iT}
\ene

\begin{equation}
{\cal D}_T=-{\chi_T\over\chi_\rho}\Gamma[(1+\Gamma_\rho)\lambda^2+2\Lambda z\lambda],\label{D_T}
\end{equation}
and $${\cal D}_r=2p\Gamma\Lambda z\lambda +6{gr\rho\over
p}(y^2-\Lambda yz).$$

The relative roles of the temperature and radius depends on the
character of perturbation and mode. As pointed out by Dziembowski,
Goode and Schou (2001, heretoforward DGS), the latter may become
dominant for f-modes, if  the magnetic perturbation is
predominantly below the region sampled by these modes. For
f-modes, to a very good accuracy, we may use
\begin{equation}
{\cal D}_r=-6{gr\rho\over p}\Lambda yz\label{D_r}
\end{equation}
In Section 9 of the present paper, we will discuss in greater
details the role of temperature and radius variation in the f- and
p-mode frequency changes.

If instead of eq.\,[\ref{delta_0_rho_T}], we use
eq.\,[\ref{delta_0_rho_s}], then we get an alternative expression
for $\Delta D_{s,0}$ which is particularly useful if the
perturbing force is localized in the deeper layers, which may be
regarded adiabatic on the eleven-year scale,
\bee
\Delta D_{s,0}&=&\int dr r^2\bigg[{\cal D}_{\rm ad}{\cal V}_0+
\nonumber\\&&
p\left({\cal D}_T{\delta S\over c_p}+{\cal D}_r{\delta r\over r}\right)\bigg],\label{D_s_02}
\ene
 where
\begin{equation}
{\cal D}_{\rm ad}=
-[\Gamma(1+\Gamma_p)+1+\Gamma_\rho]\lambda^2-2\Lambda z\lambda. \label{D_s_ad_0}
\end{equation}

\subsection{Calculations of $\Delta D_s$ for the splittings}

In the present application, it is more convenient to treat the
perturbations of the structural parameters as being Eulerian. We
consider distortions proportional to $P_{2k}$. We will  see that
within our approximation, all the angular integrals appearing in
$\Delta I$, $\Delta D_p$ and $\Delta D_g$ reduce to $Q_k$. These
factors take care of the $k$ and $m$ dependence. The property is
self-evident in the case of $D_g$. From eq.\,[\ref{D_g}] with the
use of the definitions given in eqs.\,[\ref{xi}] and [\ref{Q}], we
get
 $$\Delta
D_g=-2Q_k\int drr^3g\rho'_ky^2$$ and with eq.\,[\ref{p_k}] after
one integration by parts, and use of eq.\,[\ref{ry}], we get
\begin{equation}
\Delta D_g=4Q_k\int drr^2({\cal V}_k-{\cal H}_k)y(\lambda+\Lambda
z).
\end{equation}
The cases of $I_p$ and $D_p$ are more involved. We first note
that

$$\int_{-1}^1d(\cos\theta)\int_0^{2\pi}d\phi|\nab\Y|^2
\approx\Lambda Q_k.$$ The  approximation  assumes $\ell\gg k$,
which is not valid for low degree modes. However the terms
involving this factor are significant for such modes only in the
core, which we assume is unperturbed. Thus, for $\Delta I$ we have
approximately $$\Delta I=Q_k\int drr^4\rho'_k{\cal E}.$$ Again, we
make use of eq.\,[\ref{p_k}] and integrate by parts, and with
eqs.\,[\ref{ry}] and [\ref{rz}], to obtain approximately
\begin{equation}
\Delta I=2Q_k\int dr{r^3\over g}({\cal V}_k-{\cal
H}_k)(y\lambda+2\Lambda zy). \label{Delta_I}
\end{equation}
In calculating $\Delta D_p$, we first note that the $[...]_\phi$
term in eq.\,[\ref{Xi}] does not contribute, which follows from
the assumed axial symmetry of the perturbation. The contribution
from the $[...]_\theta$ term is nonzero, but it is  small, as may
be justified as follows. Integrating by parts over $\theta$, one
gets the $k(k+1)Q_k$ factor from the angular integral and the
whole contribution from this term is of the same order as the one
neglected above. Thus, we have \bee \Delta D_p&=&Q_k\int
drr^2\Big[\Gamma\Big(p'_k(1+\Gamma_p)+\nonumber
\\&&\rho'_k{p\Gamma_\rho\over\rho}\Big)\lambda^2
+p'_k2\Lambda({\cal E}+z\lambda)\Big]. \nonumber\ene

The quantity $\rho'_k$ is again eliminated by integration by
parts. The use is also made of eqs.\,[\ref{ry}] and [\ref{rz}]. The
result is \bee\Delta D_p&=&-Q_k\int drr^2\{{\cal
H}_k[\Gamma(1+\Gamma_p)\lambda^2+\nonumber\\&& 2\Lambda({\cal
E}+z\lambda)] +({\cal H}_k-{\cal
V}_k)\psi\},\nonumber\ene

where
\bee\psi&\equiv&\Gamma_\rho\bigg[\bigg(2-\Gamma{d\ln c^2\over d\ln
p}\bigg)\lambda^2+\nonumber\\&&\hskip-0.5cm
2\lambda\bigg(\Lambda z -{\omega^2r\over
g}y\bigg)\bigg] -\Gamma{d\Gamma_\rho\over d\ln p}\lambda^2.
\nonumber\ene
 Combining this with eq.\,[\ref{Delta_I}] in eq.\,[\ref{Delta_D_s}] ($\Delta D_g$ is negligible), we
obtain
\begin{equation}
\Delta D_{s,k}=Q_k\int drr^2({\cal D}_s^V{\cal V}_k+{\cal
D}_s^H{\cal H}_k), \label{Delta_D_sk}
\end{equation}
where we denoted
\begin{equation}
{\cal D}_s^V=-2\zeta+\psi, \label{D_s_V}
\end{equation}
$$\zeta={\omega^2r\over g}(y\lambda+2\Lambda zy),$$
and
\begin{equation}
{\cal
D}_s^H=2\zeta-\psi-\Gamma(1+\Gamma_p)\lambda^2-2\Lambda({\cal
E}+z\lambda). \label{D_s_H}
\end{equation}

\section{Frequency change to due varying turbulent pressure}

For evaluating $D_v$ according to eq.\,[\ref{D_v}], we use the
random velocity field representation given in
eq.\,[\ref{rho_v_iv_j}]. We note that
\bee
\overline{\rho|(\vv\cdot\nab)\vxi|^2}&=&\rho\sum_k\bigg({\cal
T}_k^VA^V+\nonumber\\&&\half{\cal T}_k^HA^H\bigg)P_{2k},
\ene
where $$A^V\approx \left(r{dy\over
dr}\right)^2|\Y|^2+\left(r{dz\over dr}\right)^2|\nab\Y|^2$$ and
\bee A^H&\approx&y^2|\nab\Y|^2+z^2[|(\Y)_{;\theta;\theta}|^2
\nonumber\\&&+2|(\Y)_{;\theta;\phi}|^2
+|(\Y)_{;\phi;\phi}|^2].\nonumber\ene The radial derivatives in
$A^V$ are eliminated with the help of eqs.\,[\ref{ry}] and
[\ref{rz}]. The angular integrals are evaluated by parts keeping
only derivatives of the spherical harmonics. This approximation
justifies, in particular, the replacement
$$2|(\Y)_{;\theta;\phi}|^2\rightarrow2\Re[(\Y)_{;\theta;\theta}(\Y)_{;\phi;\phi}].$$
In this way, we get the contribution to $D_v$ from the $P_{2k}$
component of the turbulent pressure,
\bee D_{v,k}&=&-Q_k\int dr
r^2\rho\bigg[{\cal T}_k^V(\lambda^2+2\Lambda z\lambda+ \nonumber\\ &&
\Lambda{\cal E}+{\cal T}_k^H{\Lambda\over2}{\cal
E}\bigg]\label{D_v_k} \ene
The contribution to $D_{s,0}$ from the
induced change in the gas pressure at constant temperature  and
radius is given by $$\Delta D_s=\int dr r^2\rho{\cal D}_{\rm
isoth}{\cal T}_0^V,$$ which follows from eqs.\,[\ref{V_k_v}] and
[\ref{D_s_01}]. Using these two expressions in eq.\, [\ref{D_o1}],
we get
\bee
(\Delta\omega)_{v,{\rm isoth}}&=&{1\over 2\omega I}\int drr^2\rho(
{\cal R}_{v,{\rm isoth}}^V{\cal T}_0^V\nonumber\\&&+{\cal R}_{v,0}^H{\cal
T}_0^H),
\ene
where $${\cal R}_{v,0}^V={\cal D}_{\rm isoth}-(\lambda^2+2\Lambda
z\lambda+\Lambda{\cal E})$$ and $${\cal
R}_{v,0}^H=-{\Lambda\over2}{\cal E}.$$ The expression for ${\cal
D}_{\rm isoth}$ is given in eq.\,[\ref{D iT}].  The complete
expressions for the ${\cal R}$'s are given in the Appendix
(eqs.\,[A1]
 and [A2]). Here, we provide only the asymptotic
forms of the ${\cal R}$'s for p-modes where $\ell|z/y|\ll1$, which
is valid sufficiently above the lower turning point,  as well as,
being the form appropriate for the f-modes. Our approximation for
p-modes is the same as that used by GMWK, and that made for
f-modes is the same as made by DGS.

For the p-modes, the leading terms are those proportional to
$\lambda^2$. If we keep only these terms, and, in addition, if we
ignore the derivatives of $\Gamma$, we find
\begin{equation}
{\cal R}_{v,{\rm
isoth}}^V\approx-\left(1+\Gamma+{\Gamma\over\chi_\rho}\right)\lambda^2,
\qquad{\cal R}_{v,0}^H\approx0. \label{R_v0_p}
\end{equation}
For the f-modes,  we have (e.g. Appendix in DGS) $y\sim \ell z$
and $\lambda\ll\ell z$ which implies
$${\cal R}_{v,{\rm isoth}}^V\approx-2\Lambda{\cal E},$$
\begin{equation}
{\cal R}_{v,0}^H=-{\Lambda\over2}{\cal E}. \label{R_v_iT_0}
\end{equation}
Thus, in both limiting cases a {\it decrease in the turbulent
pressure results in a frequency increase}. In fact, this property
is valid for all observed solar oscillations. With the help of
eqs.\,[\ref{D_s_02}] and [\ref{D_s_ad_0}], we may easily obtain
expressions for the adiabatic (ad) instead of the isothermal
kernels.

For $k>0$, we get from eqs.\,[\ref{Delta_D_sk}] and [\ref{V_k_v}]
$$\Delta D_{s,k}=Q_k\int drr^2\rho({\cal D}_s^V{\cal
T}_k^V+\half{\cal D}_s^H{\cal T}_k^H).$$ This together with the
eq.\,[\ref{D_v_k}] used in eq.\,[\ref{D_o1}] gives
\begin{equation}
(\Delta\omega)_{v,k}={Q_k\over 2I\omega}\int drr^2\rho({\cal
R}_v^V{\cal T}_k^V+{\cal R}_v^H{\cal T}_k^H),
\end{equation}
where
$${\cal R}_{v,k}^V={\cal D}_s^V-(\lambda^2+2\Lambda z\lambda+\Lambda{\cal E})$$
and
$${\cal R}_{v,k}^H=\half({\cal D}_s^H-\Lambda{\cal E}).$$
The general expressions for ${\cal D}_s^V$ and ${\cal D}_s^H$ are
given in eqs.\,[\ref{D_s_V}] and [\ref{D_s_H}], respectively.
The complete expressions for the ${\cal R}$'s are
in the Appendix (eqs.\,[A5] and [A6]). Note
that ${\cal R}_{v,k}^V$ and ${\cal R}_{v,k}^H$ are the same for
all $k>0$.

The asymptotic expressions  for p-modes are
\begin{equation}
{\cal R}_{v,k}^V\approx-2\zeta-\lambda^2,\quad {\cal
R}_{v,k}^H\approx\zeta-{\Gamma\over2}\lambda^2, \label{R_vk_V_p}
\end{equation}
where
$$\zeta\approx{\omega^2r\over g}y\lambda$$ is the highest order term in
the $\omega\rightarrow\infty$ asymptotics. However, for  solar
p-modes in the outer evanescent zone, $\zeta$ is comparable to
$\lambda^2$, and below it changes from + to - and therefore we
keep terms involving both quantities.

For the f-modes, we now have
\begin{equation}
{\cal R}_{v,k}^V\approx-3\Lambda{\cal E},\qquad {\cal
R}_{v,k}^H\approx-\half\Lambda{\cal E}. \label{R_vk_V_f}
\end{equation}
 Once we
have the ${\cal R}$ kernels, we can evaluate the $\gamma$'s
introduced in eqs.\,[\ref{D_nu_bar}] and [\ref{a}] for a specified
turbulent velocity field.  We give here an expression, which is
convenient in application to solar data \bee \gamma_{v,k}&=&\int
d\left({d_{\rm phot}\over\mbox{ 1 Mm}}\right)
\left[K_{v,k}^V\left({\delta{\cal T}_k^V\over\mbox{1
km$^2$s$^{-2}$}}\right) \right.\nonumber\\&&\left.+ K_{v,k}^H
\left({\delta{\cal T}_k^H\over\mbox{1
km$^2$s$^{-2}$}}\right)\right] \quad\mbox{$\mu$ Hz},
\label{gamma_vk} \ene where $d_{\rm phot}$ is the depth beneath
the photosphere, $$K_{v,k}^{V,H}=3.76\times10^{-8}{\mbox{1
mHz}\over\nu}x^2\tilde\rho{\cal R}_{v,k}^{V,H}.$$ The
normalization of the eigenfunctions in ${\cal R}$'s must be the
same as in $\tilde I$ used in the definition of $\gamma$.  In the
Appendix, we give exact expressions for ${\cal R}_{v,k}^{V,H}$.

\begin{figure}[]
\vskip-20mm \plotone{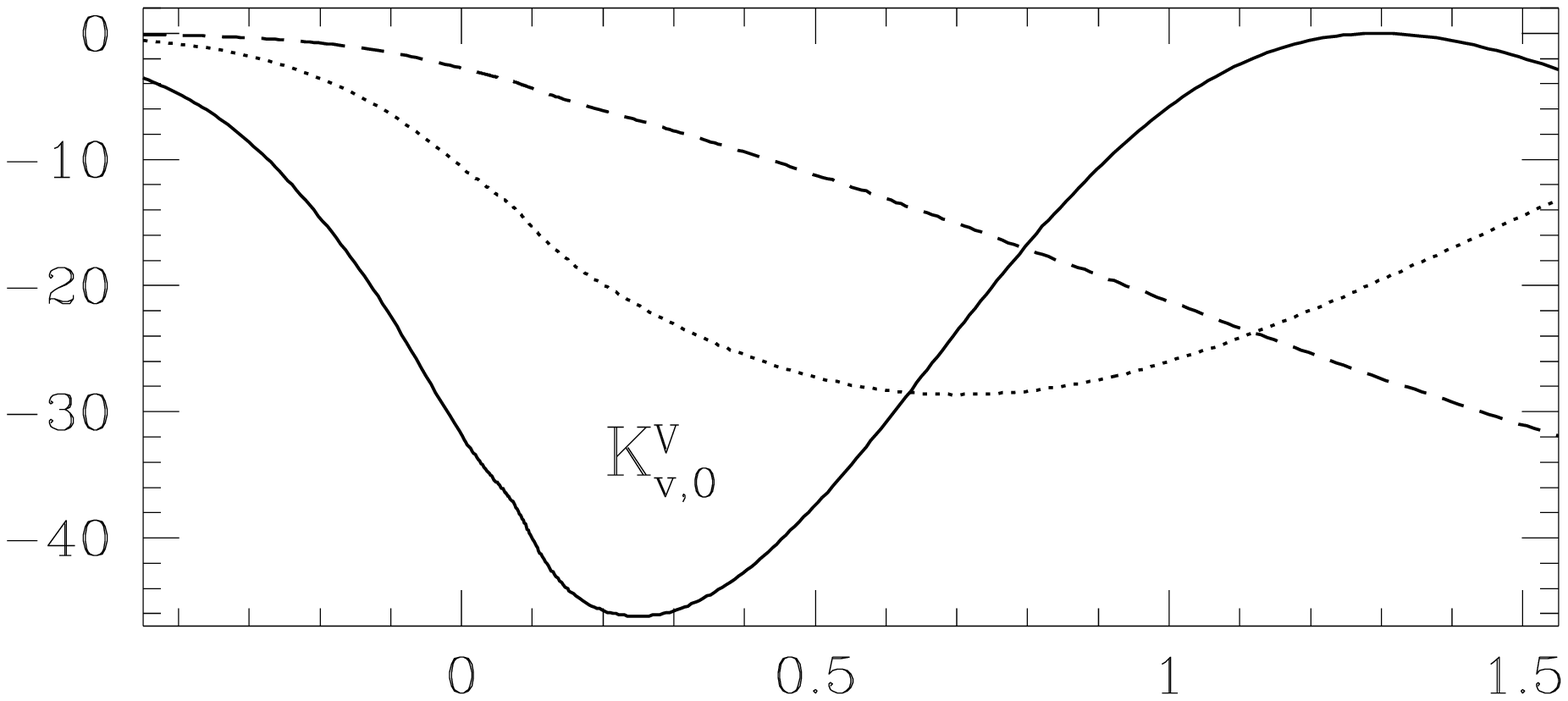} \plotone{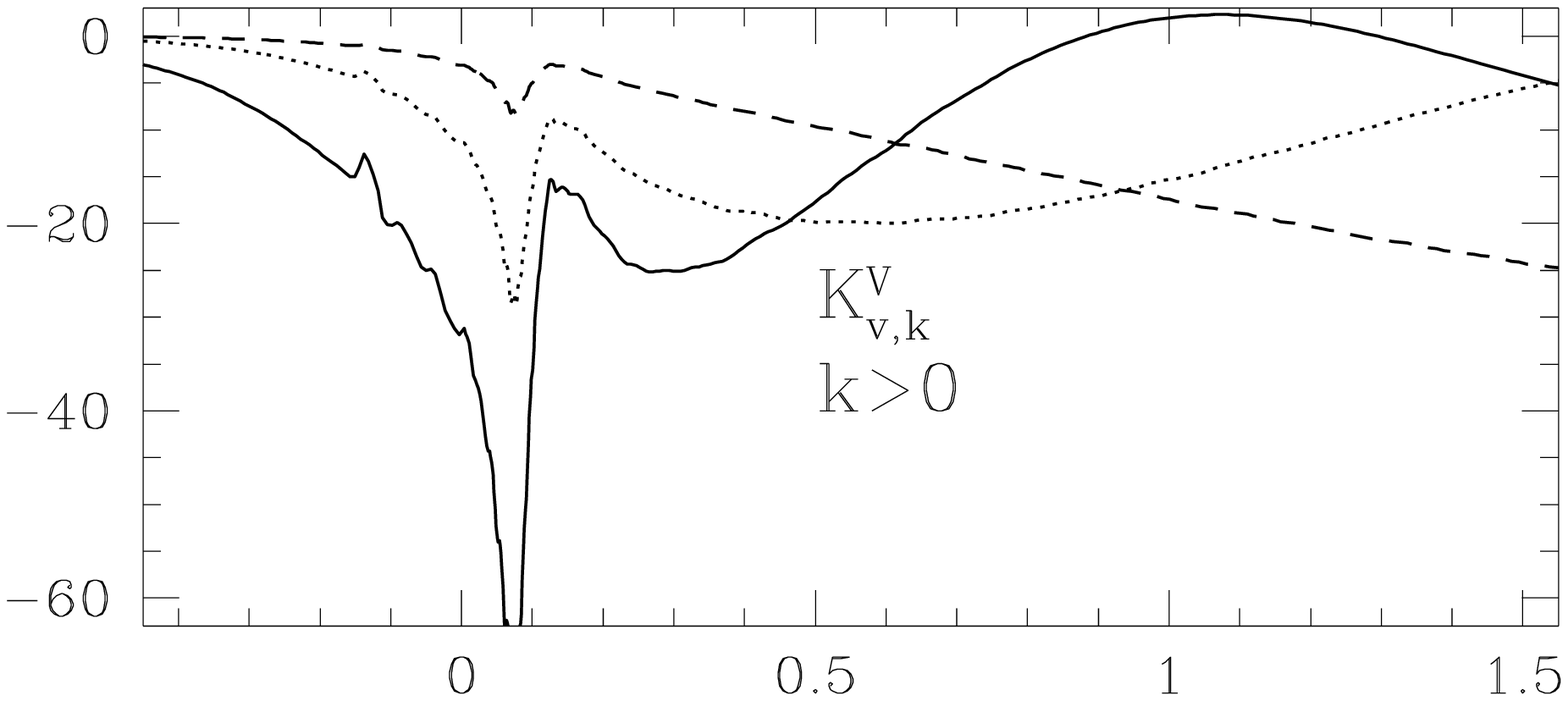} \plotone{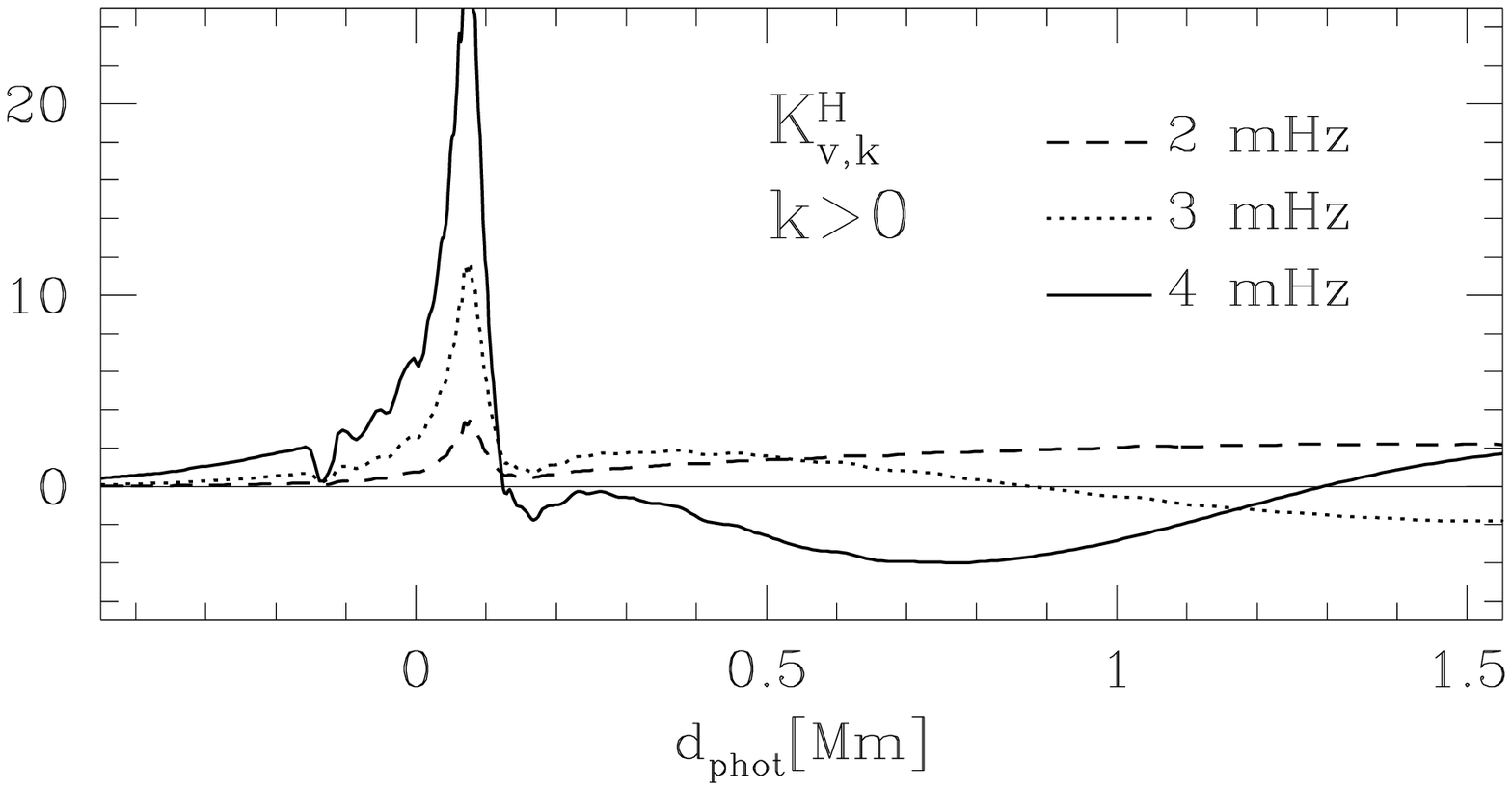}
\caption{Kernels for calculating $\gamma_{v,k}$ according to
eq.\,[\ref{gamma_vk}] at three selected frequencies plotted as
functions of the depth in outer part of the standard solar models.
The three kernels are $\ell$-independent for p-modes in this part
of the sun. The $K_{v,0}^H$ kernel is $\propto\Lambda$ and is much
smaller.}

\end{figure}

In
Fig.1, we show examples of the $K_v$  kernels that are important
for evaluating the p-mode $\gamma$'s due to the perturbation of
the turbulent pressure. Although the general trend is consistent
with the asymptotic formulae (eqs.\,[\ref{R_v0_p}] and
[\ref{R_vk_V_p}]), there are visible small-scale structures
arising from the derivatives of $\Gamma$, which we have ignored in
these two asymptotic formulae. The kernels for multiplying the
vertical component, have significantly larger absolute values and
are negative. Thus, we expect a  rise of the $\gamma$'s with a
decrease of turbulent velocities. Since the increasing magnetic
activity is expected to inhibit turbulence, the trend of the
calculated effect in $\gamma$ is consistent with observations.

With the help of Figure 1, we may roughly estimate the required
change in the mean turbulent velocity needed to account for the
measured $\gamma$'s, under the assumption that this change is the
only source of the $\gamma$'s. From numerical simulations, we know
(e.g. Abbett et al., 1997) that velocity fluctuations at the level
of 1 km/s persist over the whole layer shown in figure, with a
maximum of nearly 3 km/s at $d_{\rm phot}\approx0.1$. {\it The
frequency averaged value of $\gamma_0\approx0.3\mu Hz$ (DGS)
requires a fraction (0.2 - 0.5) of one percent decrease in the
radial component of velocity fluctuations.  The largest $\gamma_k$
($k=1$ and 3) require about one percent decrease.} Such small
changes would not be easy to detect.

\begin{figure}[ht]
\plotone{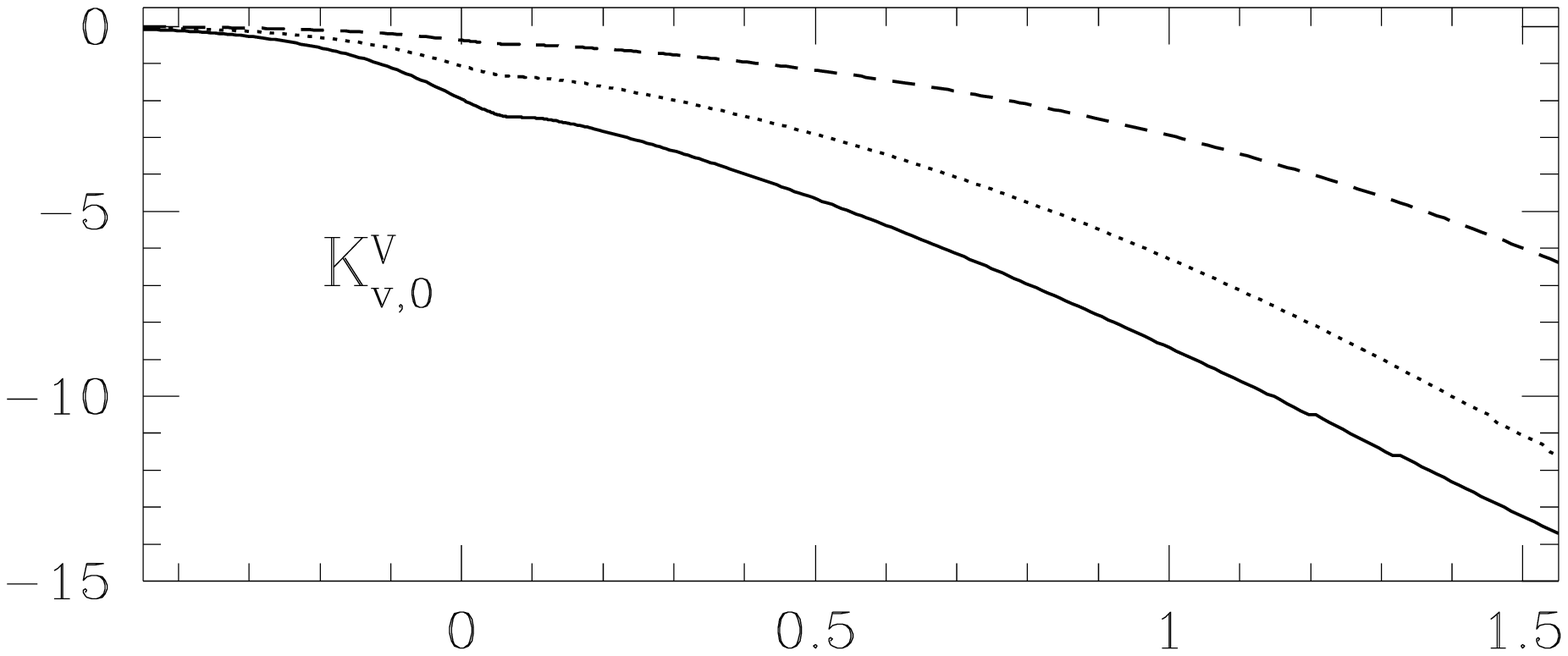} \plotone{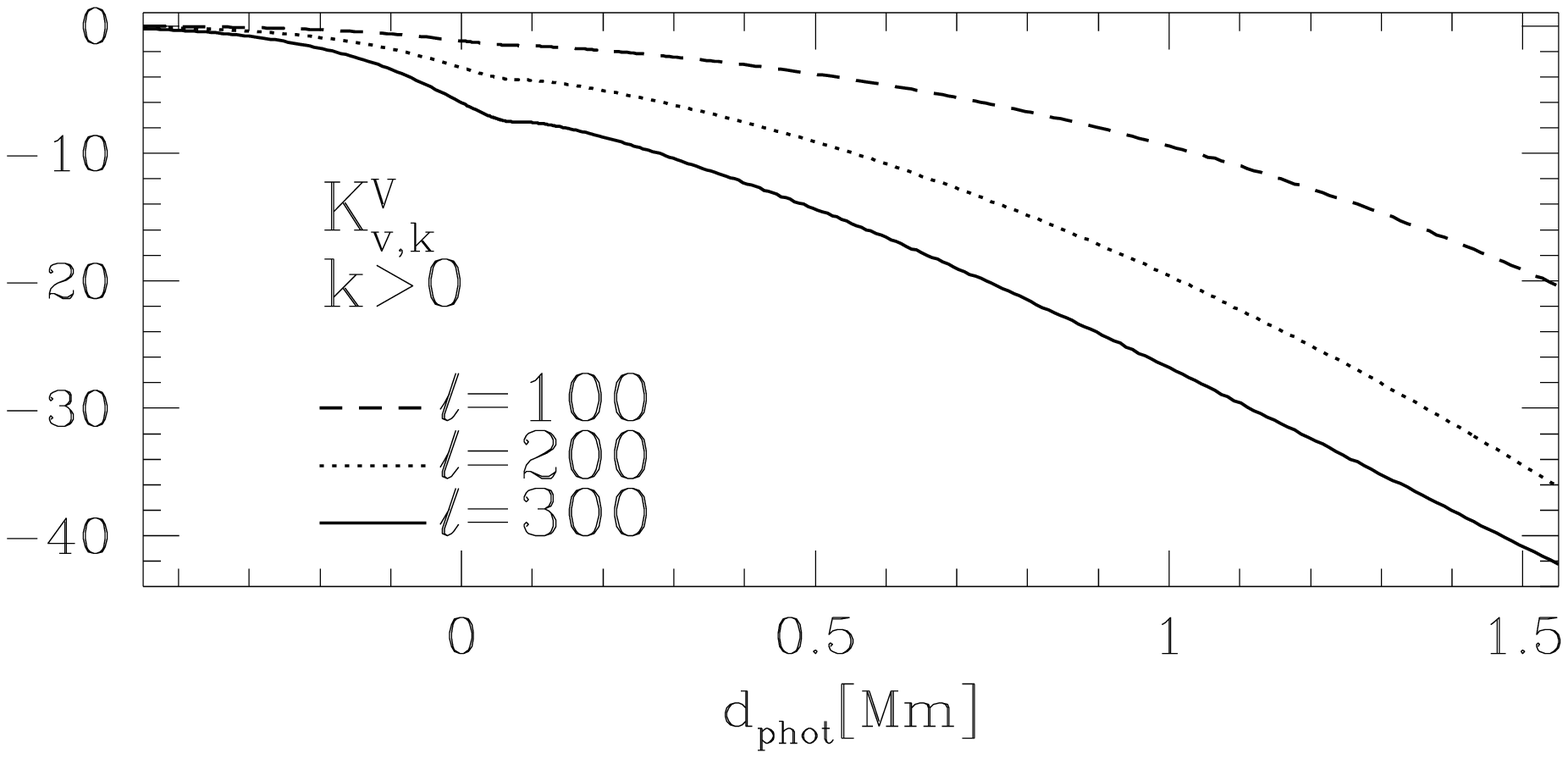} \caption{Kernels for
calculating $\gamma_{v,k}$ for f-modes at three selected
$\ell$-values.}
\end{figure}
Fig. 2 shows that the kernels for the f-modes are very different
from those for the p-modes. In this case, the asymptotic
expressions for ${\cal R}$ are quite accurate. The kernels scale
as $\Lambda/\nu\propto\ell^{1.5}$. All the kernels have similar
shapes, and they all negative. The value of the f-mode kernels are
comparable to those of the p-modes. The maximum measured values of
$\gamma_0$ for f-modes are about twice that for p-modes, but the
errors are large. The uncertainty for $k>0$ is even higher.

More detailed analyses of the $\gamma_k(\nu)$ dependence
reflecting the kernels frequency dependence seen in Figs. 1 and 2
are needed to say more about the nature of the required change in
the turbulent velocity. However, at this stage we may already
conclude that this change must be regarded as important, perhaps
the dominant contributor to solar p-mode and f-mode frequency
changes over the activity cycle. The sign of the observed changes
agrees with the expected inhibiting effect of the field on
convection.

\section{Frequency change due to varying small-scale, near-surface magnetic field}

With the random field being described by a single
$P_{2k}$-component (see eq.\,[\ref{B_iB_j}]), the three terms in
the integrand of eq.\,[\ref{D_M}] are transformed as follows
\bee&&\overline{|(\vB\cdot\nab)\vxi|^2}\rightarrow
\nonumber\\&&\left[{\cal M}_k^V(\lambda^2+2\Lambda
z\lambda+\Lambda{\cal E})+ {\cal M}_k^H{\Lambda\over2}{\cal
E}\right]Q_k,\nonumber\ene

\bee&&-2\overline{\di\vxi^*\vB\cdot(\vB\cdot\nab)\vxi}\rightarrow
\nonumber\\&&-[2{\cal M}_k^V\lambda(\lambda+\Lambda z) -{\cal
M}_k^H\Lambda z\lambda]Q_k,\nonumber\ene and
\bee&&\half\overline{|\vB|^2}(\Xi+|\di\vxi|^2)\rightarrow
\nonumber\\&&({\cal M}_k^V+{\cal
M}_k^H)[\lambda^2+\Lambda(z\lambda+{\cal E})]Q_k.\nonumber\ene The
transformations use integration by parts over the surface, our
approximations regarding the eigenfunctions, and the angular
dependence of the averaged fields. Note that the first term is
fully analogous to the integrand in $D_v$ that was considered in
the previous section.

With the above expression, we get from eq.\,[\ref{D_M}] \bee
D_{M,k}&=&{Q_k\over4\pi}\int drr^2\{\delta{\cal
M}_k^V\Lambda(z\lambda+2{\cal E})+ \nonumber\\&&\delta{\cal
M}_k^H[\lambda^2+\Lambda(2z\lambda+1.5{\cal E})]\}. \nonumber\ene

For the spherically symmetric part of $\Delta D_s$, we use
eq.\,[\ref{D_s_01}], ignoring here again the temperature and
radius changes. This combined with eq.\,[\ref{V_k_B}], yields
\begin{equation}
\Delta D_{\rm isoth}={1\over8\pi}\int dr r^2{\cal D}_{\rm ad}
 ({\cal M}_0^H-{\cal M_0}^V). \label{Delta_D_s_iT}
\end{equation}
Using last two expressions in eq.\,[\ref{D_o1}], we get \bee
(\Delta\omega)_{M,{\rm isoth}}&=&{1\over16\pi\omega I} \int
drr^2({\cal R}_{M,{\rm isoth}}^V{\cal M}_0^V \nonumber\\&&+{\cal
R}_{M,{\rm isoth}}^V{\cal M}_0^H), \label{Delta_omega_M_0} \ene
where $${\cal R}_{M,{\rm isoth}}^V=2\Lambda(z\lambda+2{\cal E})-
{\cal D}_{\rm isoth}$$ and $${\cal R}_{M,{\rm isoth}}^H=2\lambda^2+\Lambda(4z\lambda+3{\cal E})+
 {\cal D}_{\rm isoth},$$ with ${\cal D}_{s,{\rm isoth}}$ is given in eq.\,[\ref{D iT}].
 The complete expressions for the ${\cal R}$'s are in the
Appendix (eqs.\,[A3] and [A4]).
 The asymptotic expressions for the p-modes are
 $${\cal R}_{M,{\rm isoth}}^V\approx\left(\Gamma+{\Gamma\over\chi_\rho}\right)\lambda^2$$
and
\begin{equation}
{\cal R}_{M,{\rm isoth}}^H\approx\left(2-\Gamma-{\Gamma\over\chi_\rho}\right)\lambda^2.
\end{equation}
The adiabatic kernels,
 equivalent to those found by GMWK, are obtained by replacing $\chi_\rho$ with $\Gamma^{-1}$.
In both cases {\it the equations
imply that an increase in the vertical component leads to a
decrease in the frequencies, while the opposite is true for the
horizontal component}. The sign of frequency shift due to the
horizontal field is opposite to what one might have naively expected
because the dominant effect of such field arises through the
perturbation of the equilibrium structure (the $D_s$ term) and not
by the direct effect of the field on oscillations (the $D_M$
term). The former term is negative because the horizontal field
causes a local expansion, hence an increase of the sound
propagation time. The vertical field has an opposite effect.
An isotropic (${\cal M}_0^H=2{\cal M}_0^V$) field increase implies a net frequency
increase but the required increase to account for the observed frequency changes
is large.

For the f-modes, the ${\cal D}_{s,{\rm isoth}}$ may be neglected
and we have
\begin{equation}
{\cal R}_{M,{\rm isoth}}^V\approx{\cal R}_{M,{\rm isoth}}^H\approx{4\over3}\Lambda{\cal E}.
\end{equation}
Thus, {\it an increase of either component of the magnetic
field implies a frequency increase for f-modes}.

For $k>0$ we have from eqs.\,[\ref{Delta_D_sk}] and [\ref{V_k_B}]
\bee \Delta D_{s,k}&=&{Q_k\over8\pi}\int drr^2[({\cal D}_s^H-{\cal D}_s^V){\cal M}_k^V
\nonumber\\&&+{\cal D}_s^V{\cal M}_k^H)]\nonumber\ene
 and
\bee
(\Delta\omega)_{M,k}&=&{Q_k\over16\pi\omega I}\int drr^2({\cal
R}_M^V{\cal M}_k^V\nonumber\\&&+{\cal R}_M^H{\cal M}_k^H), \label{Delta_omega_Mk}
\ene
where
$${\cal R}_{M,k}^V={\cal D}_s^H-{\cal D}_s^V+2\Lambda(z\lambda+2{\cal E})$$
and
$${\cal R}_{M,k}^H={\cal D}_s^V+2\lambda^2+\Lambda(4z\lambda+3{\cal E}).$$
Expressions for ${\cal D}_s^V$ and ${\cal D}_s^H$ are given in
eqs.\,[\ref{D_s_V}] and [\ref{D_s_H}], respectively. The complete
expressions for the ${\cal R}$'s are in the Appendix (eqs.\,[A7]
and [A8]). Again kernels ${\cal R}_{M,k}$ are the same for all
$k>0$. The The asymptotic expressions  for p-modes are
\begin{equation}
{\cal R}_{M,k}^V\approx4\zeta-\Gamma\lambda^2,\qquad {\cal
R}_{M,k}^H\approx-2\zeta+2\lambda^2.
\end{equation}
For the f-modes, we now have
\begin{equation}
{\cal R}_{M,k}^V\approx6\Lambda{\cal E},\qquad {\cal
R}_{M,k}^H\approx\Lambda{\cal E}.
\end{equation}
Now we have for $\gamma$'s
\bee
\gamma_{M,k}&=&\int d\left({d_{\rm phot}\over\mbox{ 1 Mm}}\right)
\left[K_{M,k}^V\left({\delta{\cal M}_k^V\over\mbox{1
kG}}\right)^2\right.\nonumber\\&&\left.+ K_{M,k}^H \left({\delta{\cal M}_k^H\over\mbox{1
kG}}\right)^2\right] \quad\mbox{$\mu$ Hz}, \label{gamma_M_k}
\ene
where
$$K_{M,k}^{V,H}=1.06\times10^{-13}{\mbox{1 mHz}\over\nu}x^2{\cal R}_{M,k}^{V,H}.$$
\begin{figure}[ht]
\plotone{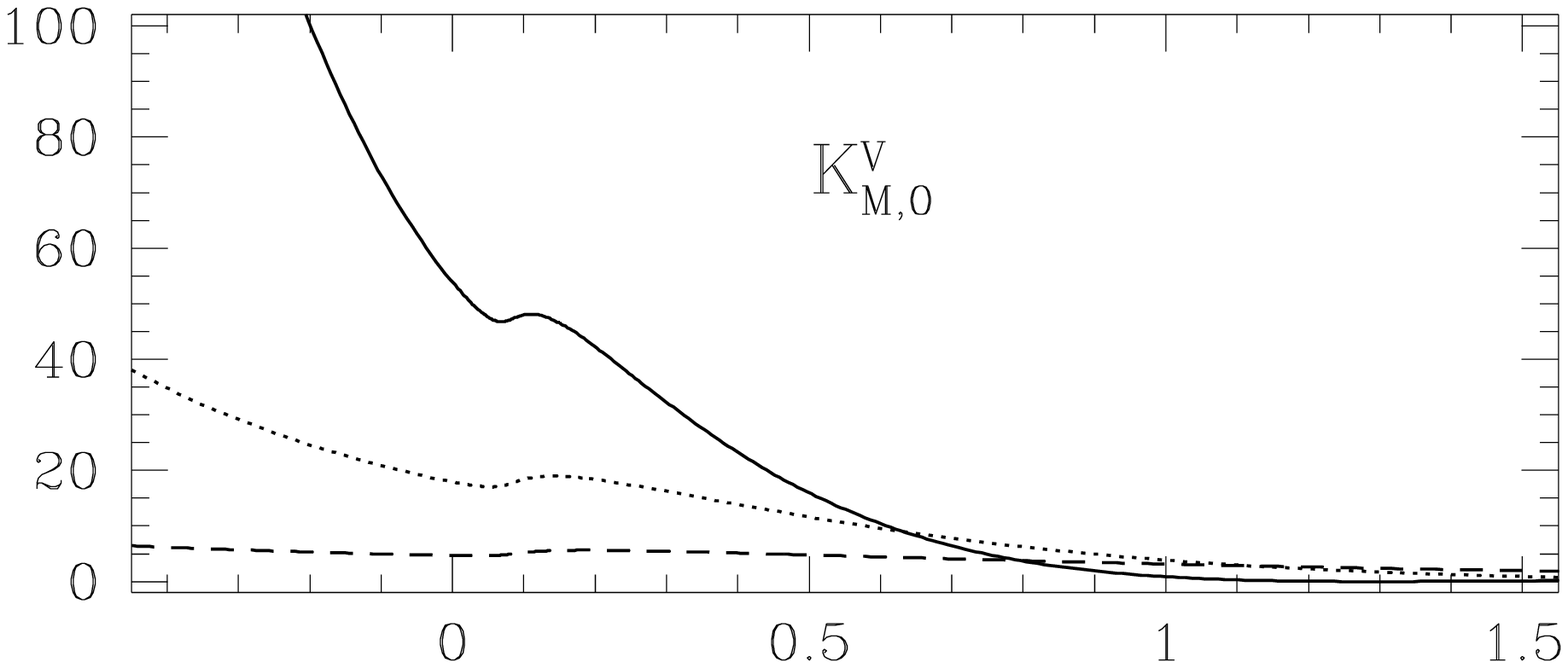} \plotone{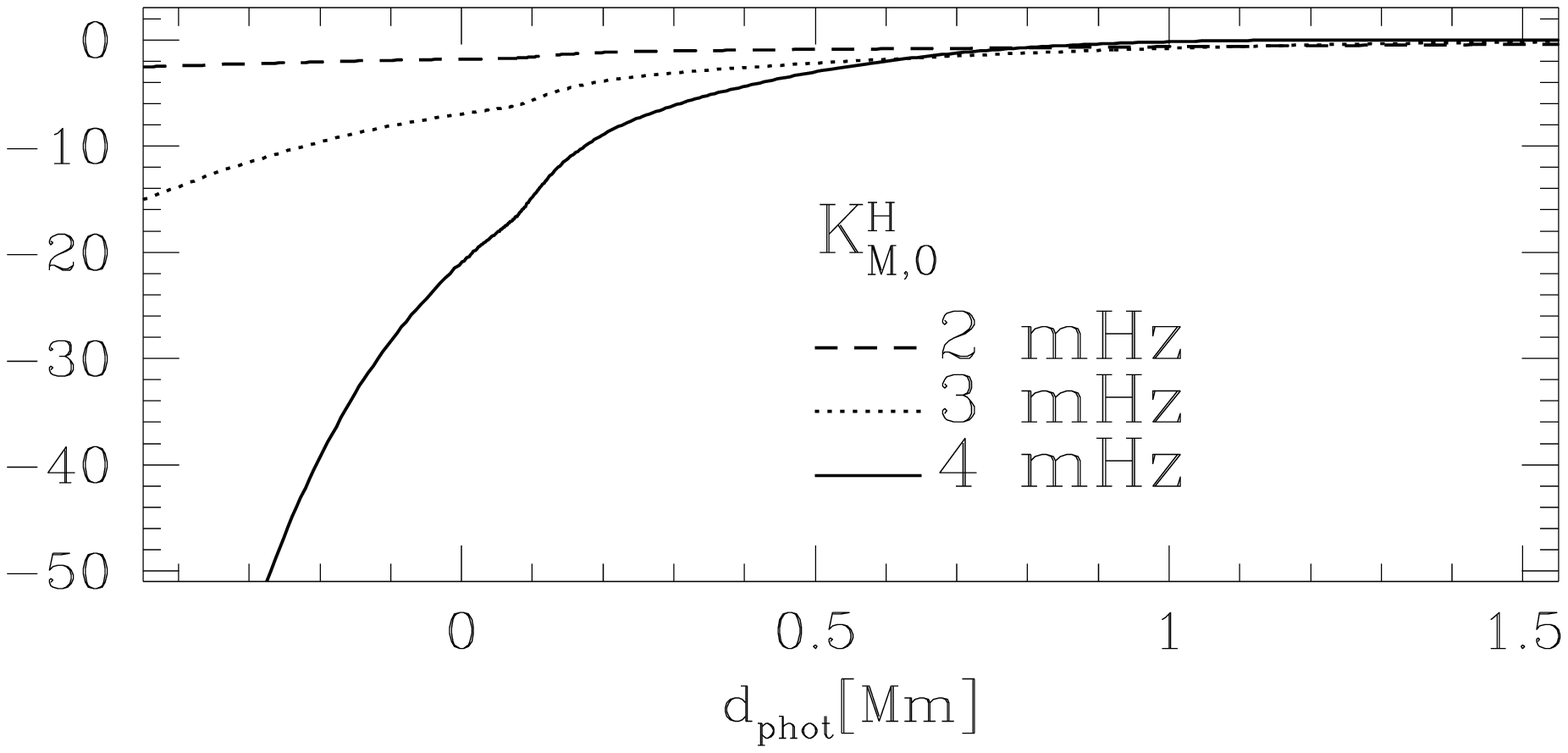} \caption{Kernels for
calculating  centroid frequency shifts due to a small-scale
magnetic for p-modes at the three selected frequencies.}
\end{figure}
\begin{figure}[ht]
\plotone{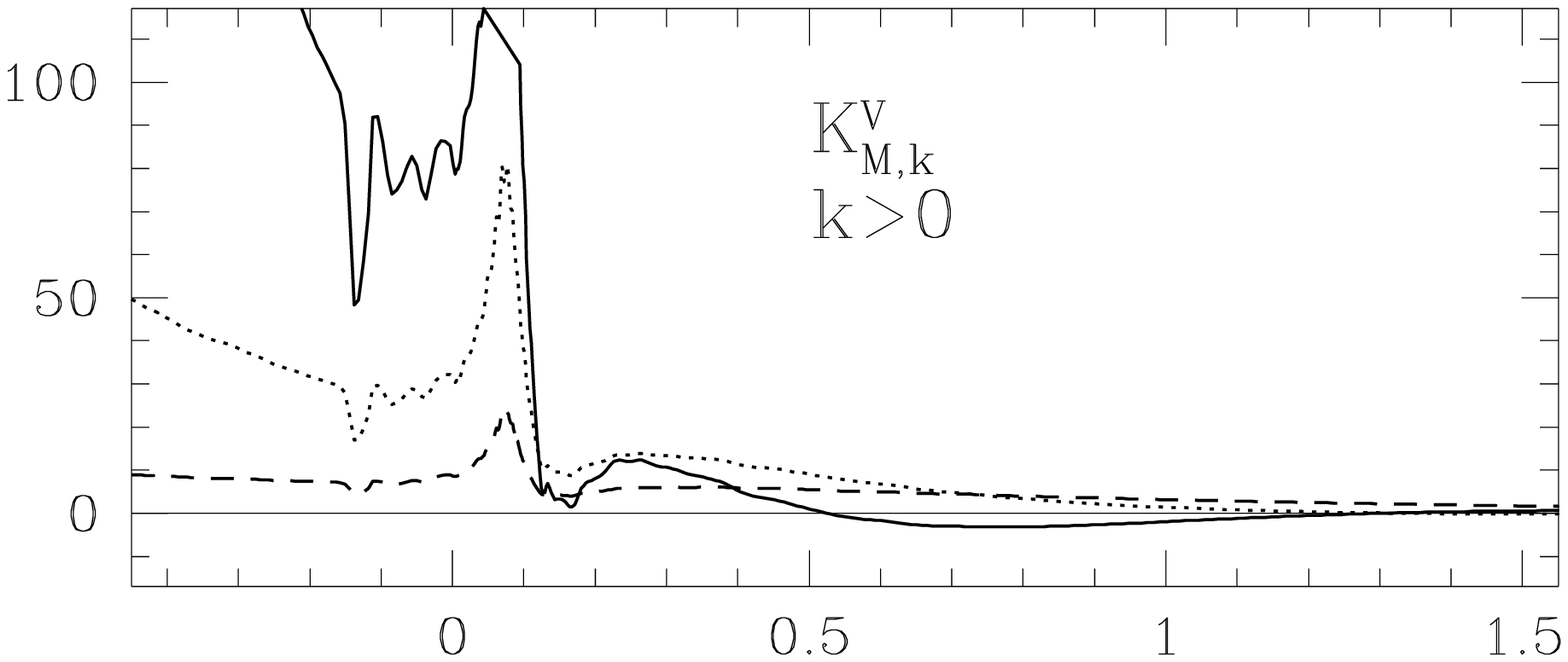} \plotone{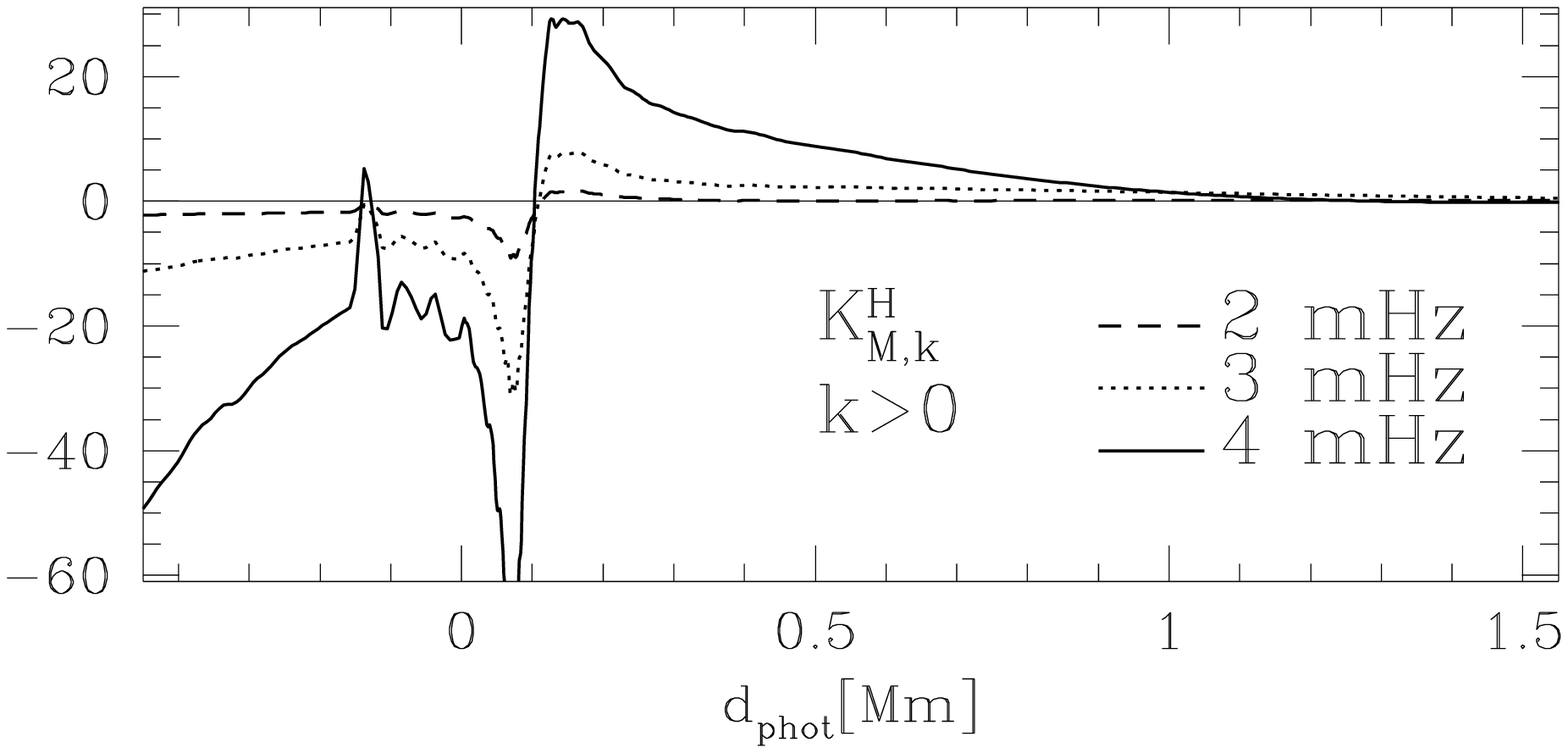} \caption{The same as Fig.3 but
for calculating the even-$a$ part of the frequency splitting }
\end{figure}
In Figs. 3 and 4, we show  kernels  for calculating $\gamma_M$
according to eq.[\ref{gamma_M_k}]. Note the strong sensitivity to
the frequency, which emphasizes the probing power of the
$\gamma(\nu)$ dependence. Further, note that {\it the kernels
imply that an increase in the radial field in outer layers will
lead to an increase in the mean frequency, while that of the
horizontal field has the opposite effect.} The growth of the
vertical field also leads to an increase of $\gamma$'s at $k>0$,
but the effect of the horizontal field growth is impossible to
predict as it depends a lot on the depth where it takes place.
Also in the case of magnetic fields the kernels for f-modes differ
significantly from those for p-modes as we may see comparing Fig.
5 with those in Figs. 3 and 4. For f-modes, the effect of the
horizontal components of the field is similar to the vertical ones
but smaller.

\begin{figure}[ht]
\plotone{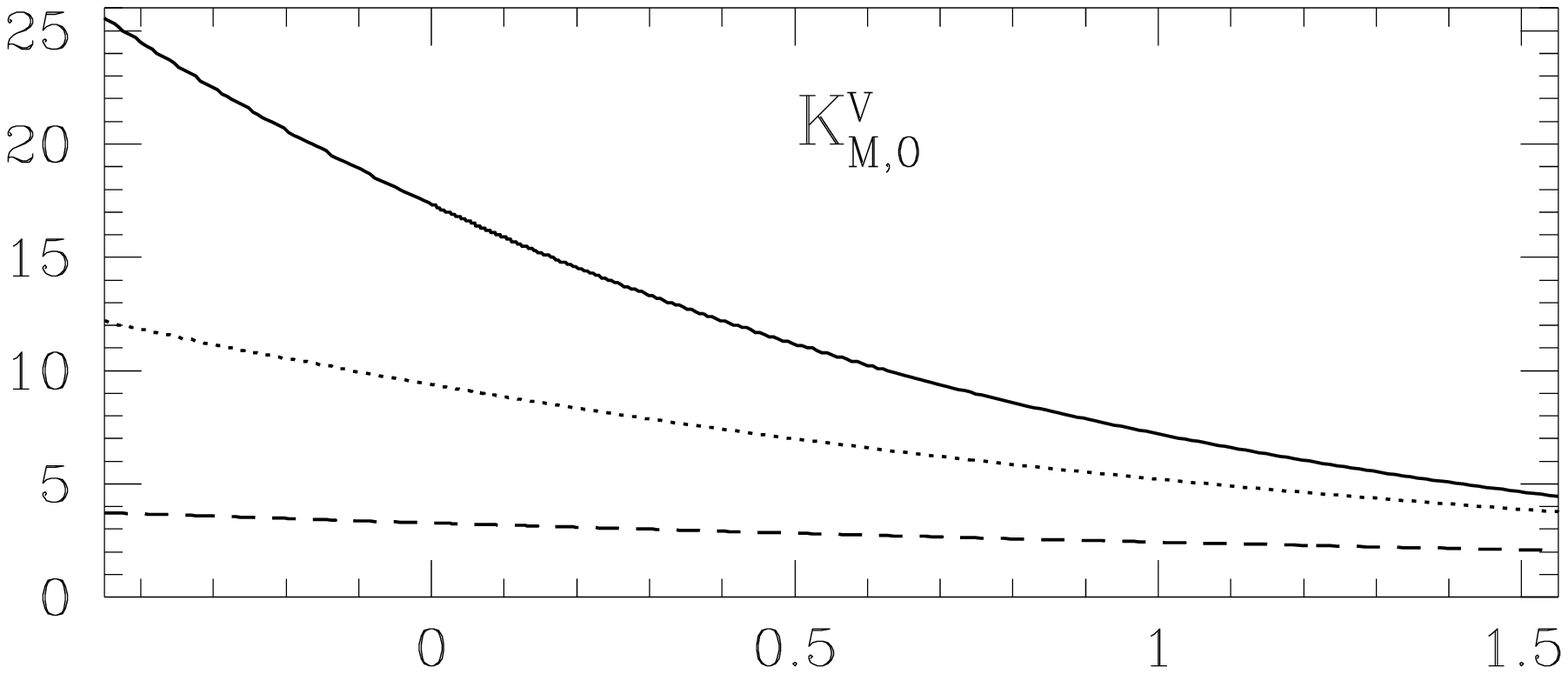} \plotone{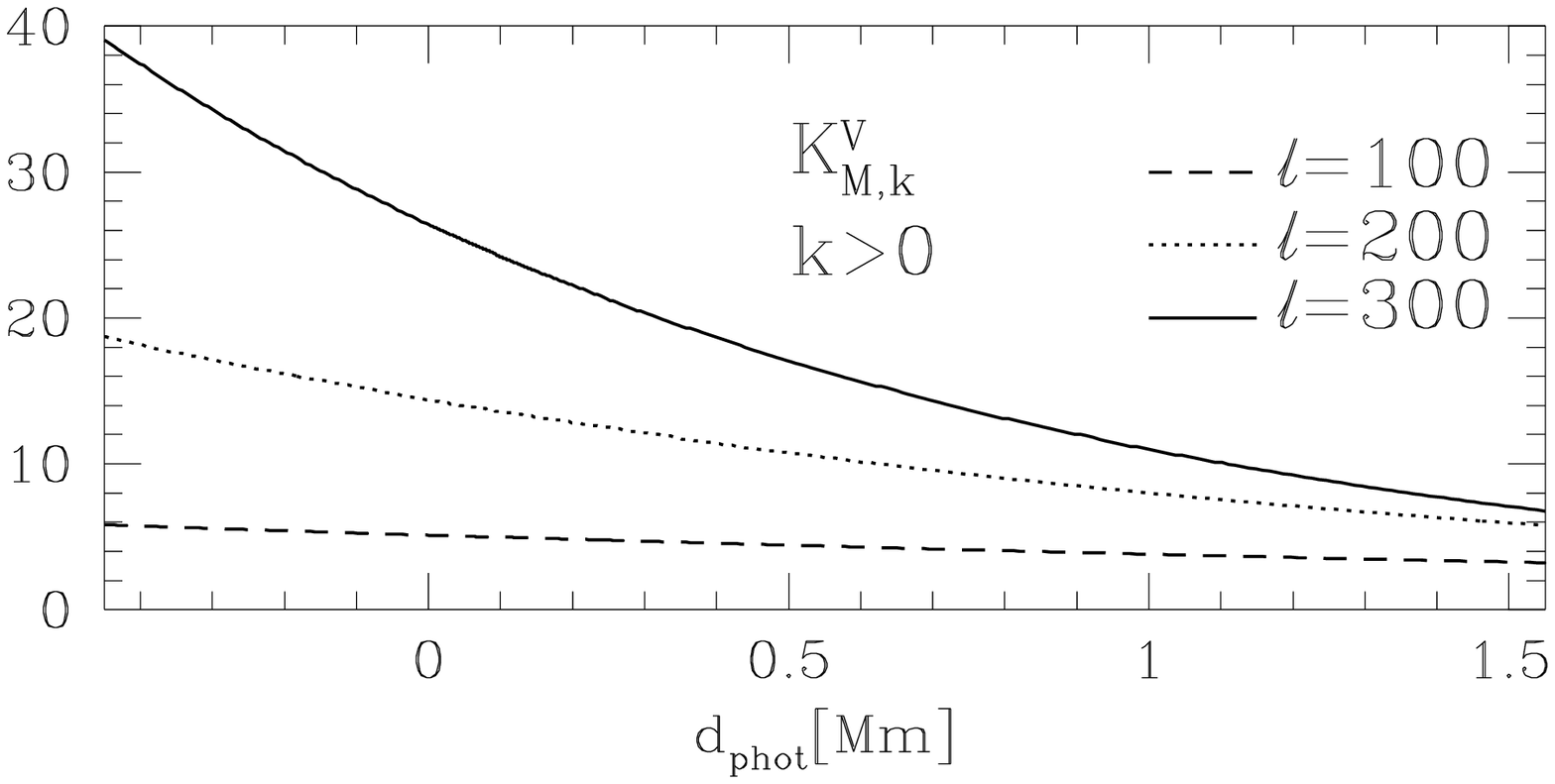} \caption{Kernels for
calculating the centroid shifts (upper panel) and splittings due
to vertical components of the field for f-mode modes at three
selected degrees. The kernels due to the horizontal components
have similar shapes but smaller (factor $\approx3$ at $k=0$ and
$\approx6$ at $k>0$).}
\end{figure}
Again, we may use the plots shown in these figures to assess the
required magnetic field changes needed to account for the measured
$\gamma$'s. Let us first consider $\gamma_0$. For p-modes, the
minimum requirement for the field increase is obtained, if we
assume that only radial component increases and it is $\sim 100$ G
(DGS).  The number rises to above 200 G if we assume an isotropic
field increase (GMWK, DGS).  This latter value is unacceptably
high. Also, there is a higher requirement to account for f-mode
$\gamma_0$'s. Though the observational accuracy is poorer than in
case of p-modes, this may be regarded as a piece of evidence
against the direct effect of a changing magnetic field as the sole
cause of the frequency changes. Also accounting for the even-$a$
coefficients sets more stringent requirements on the near-surface
magnetic field, which may be difficult to reconcile with the
measurements.

\section{Temperature and radius variation}

GMWK were first to considered the role of temperature variations
in the p-mode frequency changes.  They correctly observed that the
temperature decrease at constant pressure results in frequency
decrease because the effect of local expansion exceeds that of
sound speed increase. However, they excluded the effect of
temperature change as a primary source of the measured frequency
changes. Here we reconsider the effect using our formulation
presented in Section 5.1. With eqs.\,[\ref{D_o2}],[\ref{D_o1}],
[\ref{D_s_01}], and [\ref{D_T}] we obtain the following expression
for the temperature contribution to $\gamma_0$.
\bee
\gamma_{T,0}&=&\int d\left({d_{\rm phot}\over\mbox{ 1
Mm}}\right)K_T\nonumber\\&&{\delta T\over T}\quad \mu{\rm Hz}, \ene
where
$$K_T=-0.72{\nu\over\mbox{1 mHz}}\left({c\over\omega r}\right)^2x^4\tilde\rho
{\chi_T\over\chi_\rho}{\cal R}_T$$
and
$${\cal R}_T=(1+\Gamma_\rho)\lambda^2+2\Lambda z\lambda.$$
For p-modes the first term is dominant in ${\cal R}_T$. Except for
our taking into account the derivative of $\Gamma$, it is the same
as in GMWK. For f-modes the second term is much greater but the
entire kernels are much smaller than for p-modes, as we may see in
Fig. 6. Thus, we will consider the effect of temperature only for
the p-modes. With the plots in the upper panel, we may estimate
that the fractional temperature increase in the outer layers at a
$10^{-4}$ level implies decrease of $\gamma_0$ at a $10^{-1}
\mu$Hz level, which is significant. The question arises whether
such temperature changes during the solar cycle are feasible.
\begin{figure}[ht]
\plotone{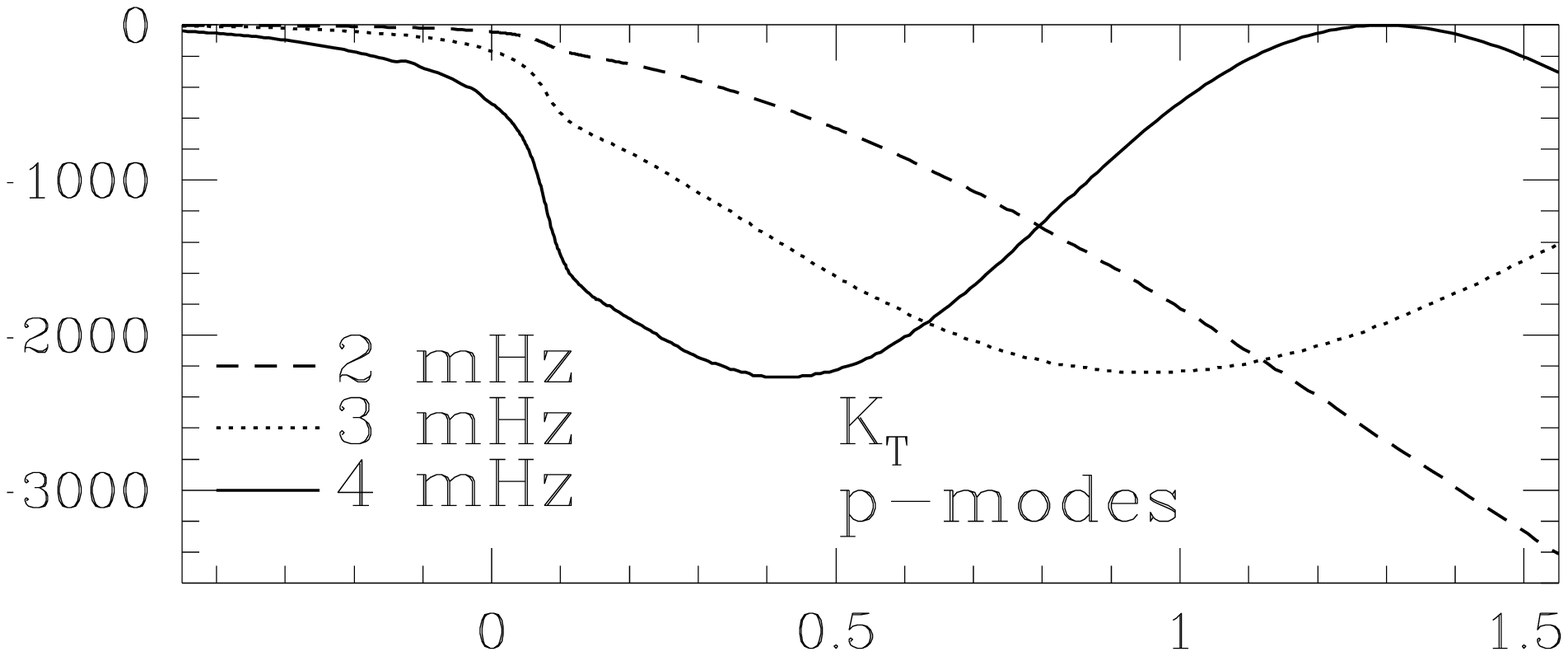} \plotone{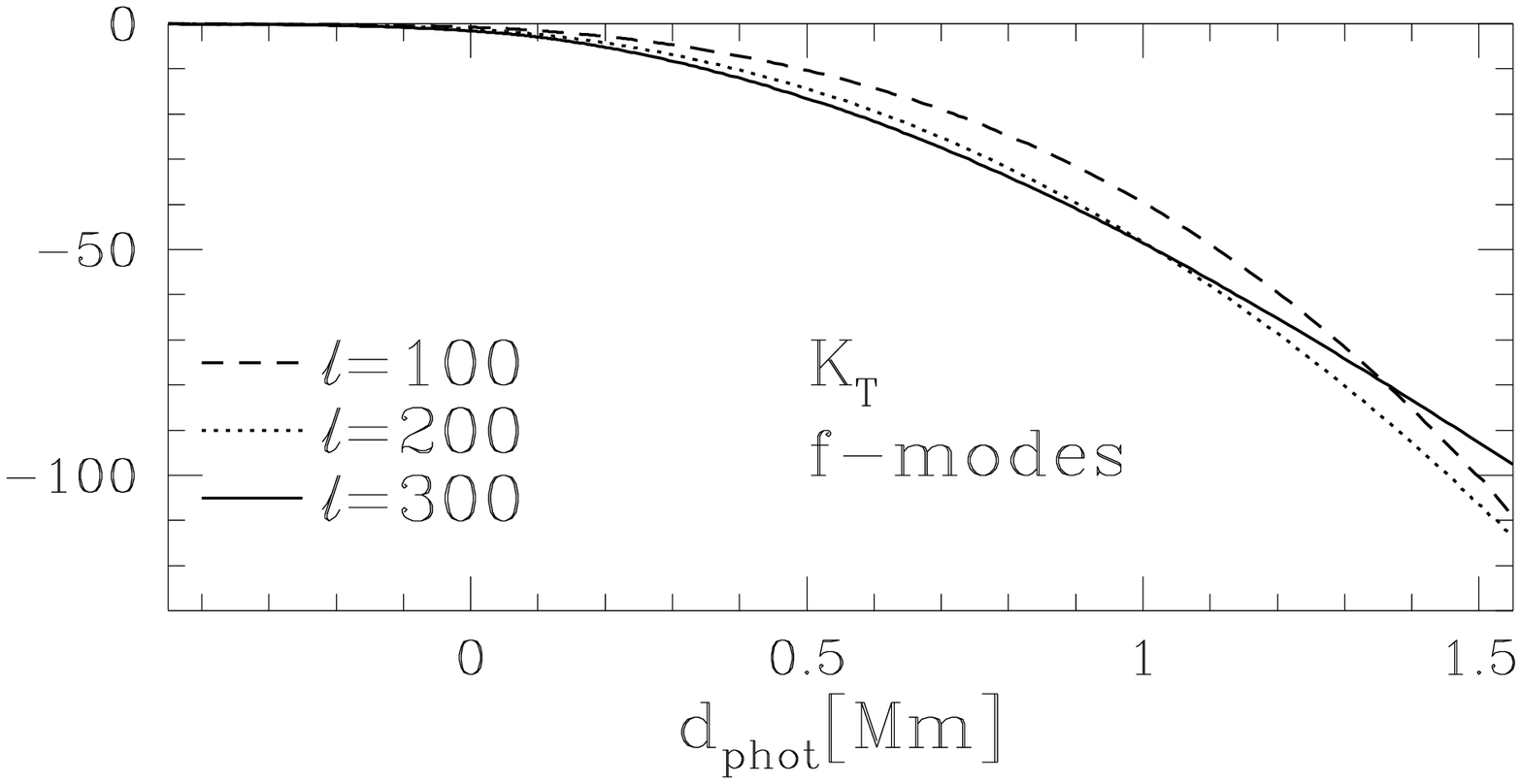} \caption{Kernels for
calculating the frequency shifts due to temperature increase for
p-modes at the selected frequencies (upper panel) and f-mode modes
at the selected degrees (lower panel).}
\end{figure}

Temperature variation correlated with magnetic variations are
expected. However even the sign of it is a matter of debate. Gray
\& Livingston (1997) put forward evidence that there an increase
of $T_{\rm eff}$ between the activity minimum and maximum by some
1.5K, that is  $\delta T_{\rm eff}/T_{\rm
eff}\approx2.6\times10^{-4}$, which would account for the observed
variation in the solar constant. Since the optical depth increases
with temperature increases,  $\delta T$ at $d_{\rm phot}=0$ is
somewhat greater than $\delta T_{\rm eff}$. An estimate using the
Eddington approximation yields $\delta T(d_{\rm phot})=1.4\delta
T_{\rm eff}$. The result of Gray \& Livingston is not generally
accepted. Spruit(1991) argues that the dominant effect of the
magnetic field on temperature is through inhibition of convection
and hence it implies cooler layer outer layers at high activity.
If this indeed the case, then the induced temperature variation
contribute to frequency increase. Spruit (1991) explains the
irradiance increase correlated with the activity as a result of an
increased corregation of the photosphere.

In a crude manner, the expected temperature change may be linked
to the change in the turbulent velocity. In Section 6, we found
that $(2-5)\times10^{-3}$ change in turbulent velocity suffices to
explain the maximum value of $\gamma_0\approx0.3 \mu$Hz.  Our aim
is to estimate the values of $\delta T/T$ in the subphotospheric
layer extending down to (say) 1 Mm associated with such a change
in the velocity. To this aim, we rely on the mixing-length
approximation (MLT) and we mimic the inhibiting effect of the
field by varying the MLT parameter $\alpha$. While perturbing
$\alpha$, we keep both $R$ and entropy constant in the adiabatic
part of the convective zone. Adopting $\delta
v/v=-3\times10^{-3}$, we find $\delta T/T=-1\times10^{-3},
-2\times10^{-4},\mbox{ and} -1\times10^{-4}$ at $d_{\rm phot}=0$,
0.5, and 1 Mm, respectively. The implied 4 K decrease of $T_{\rm
phot}$ between solar minimum and maximum seems unacceptably large.
This, however, should not be regarded as a case against changes in
turbulent velocities being the primary source of the frequency
changes because our treatment of energy transport was very crude
indeed. Rather, we want to emphasize here that {\it temperature
changes in the subphotospheric layers must be considered as
significant contributor to the observed frequency changes over the
solar cycle}.

The aspherical part of the temperature perturbation is fixed by
the condition of mechanical equilibrium.  Eq.\,[\ref{delta_0_T}]
expresses $\delta_k T$ in terms of the expansion coefficients
${\cal V}_k$ and ${\cal H}_k$, which in turn are linked to the
expansion coefficients for turbulent pressure (eq.\,[\ref{V_k_v}])
and magnetic field (eq.\,[\ref{V_k_B}]). We may see that any
inference on temperature depends on the derivative of the
perturbing force, and thus requires a detailed analysis of the
$\gamma(\nu)$ dependence. Currently available data are probably
not accurate enough for this.

In contrast to the effect temperature, which may only be important
for p-modes, the effect of radius perturbation is likely to play a
role only in f-mode frequency changes. Considering in
eq.\,[\ref{D_s_01}] only the effect of radius perturbation and
adopting the approximation $2\Lambda yz\approx{\ell\cal E}$, which
is valid for f-modes, we get from eq.\,[\ref{D_o1}]
$$\left({\delta\nu\over\nu}\right)_{r,0}=-{3\ell\over2\omega^2I}\int
dI{g\over r}{\delta r\over r}.$$ This expression was used by DGS,
who argued that the part of the frequency increase which is
proportional to $\nu$ may be explained by the part of $\delta r$
which is common to all modes in the f-mode set. Their set
contained modes with $\ell$'s from 137 to 300. The common part
must  originate below the outer part of the sun sampled by all
these modes, that is below radius $r=0.988R_\odot$. They argued
that its likely cause is an increase in the radial component of
the magnetic field by few kG. It is ironic that the best evidence
for deep seated magnetic field changes may come from modes which
do not directly probe the region where the field is located.
Unfortunately, what we get with these modes is only an integral
constraint on the field. Of course, it would be advantageous to
have a direct probe for the deep seated field.

\section{Frequency perturbation due the horizontal field
in deep layers}

It has been argued (see, e.g., D'Silva \& Howard 1993) that a
horizontal field of $B\sim 10^5$ G is present in the region near
the base of the convective envelope. Seismic evidence for the
presence of such a field is still controversial.

First, we consider a large-scale toroidal field of the form given
in eq.\,[\ref{B_t}], and truncated at $j=2$. The consecutive terms
in $D_M$(see eq.\,[\ref{D_M}]) are calculated under the same
approximation as used in \S3. The three integrands become
$$|(\vB\cdot\nab)\vxi|^2=Wm^2(y^2|\Y|^2+z^2|\nab_H\Y|^2),$$
$$-2\di\vxi^*\vB\cdot(\vB\cdot\nab)\vxi=-2Wm^2|\Y|^2z\lambda,$$
and \bee
\half|\vB|^2(\Xi+|\di\vxi|^2)&=&W(1-\mu^2)|\Y|^2\nonumber\\&&
[\lambda^2+\Lambda({\cal E}+z\lambda)],\nonumber \ene
where we denoted
$$W\equiv{3\over4}B^1_{t,2}+{15\over4}B^2_{t,2}\mu^2+
{3\over2}\sqrt{5}B^1_{t,2}B^2_{t,2}\mu. $$ Note that the last term
in $W$ cancels out upon integration. Calculating the surface
integrals first two terms in $D_M$, we rely on the following
recursion relation (DG 91). \bee q_k&=&{2k-1\over
k(4\Lambda+1-4k^2)}\{q_{k-1}[2\Lambda-
\nonumber\\&&\hspace{-0.8cm}(2k-1)^2-2m^2+q_{k-2}(2k-3)(k-1)\},\nonumber \ene where
$$q_k=\int_{-1}^1d\mu\int_0^{2\pi}d\phi\mu^{2k}|\Y|^2.$$ With this
relation, assuming
 $\Lambda\gg1$ and using explicit expressions for $P_{2k}$
the surface integral becomes \bee
m^2\int_{-1}^1d\mu\int_0^{2\pi}d\phi\mu^2|\Y|^2&\approx&\nonumber\\
\Lambda\left({1\over15}-{2\over21}Q_1-{32\over105}Q_2\right).\nonumber
\ene We could assume $\Lambda\gg1$ because for low degree modes
the third terms dominates. The surface integral in this term is
easily expressible in terms of $Q_1$ and $Q_2$. Combining all the
three terms in eq.\,[\ref{D_M}], we obtain \bee
D_M&=&{1\over4\pi}\int
drr^2\bigg\{B_{t,1}^2\bigg[{1\over4}(2\lambda^2+3\Lambda{\cal E}
) \nonumber\\&&\hspace{-0.8cm}-{Q_1\over2}(\lambda^2+3\Lambda{\cal E}-3\Lambda
z\lambda)\bigg]+
B_{t,2}^2\nonumber\\&&\hspace{-0.8cm}\bigg[{1\over4}(2\lambda^2+3\Lambda{\cal
E})+{5Q_1\over14}(\lambda^2+3\Lambda z\lambda)
\nonumber\\&&\hspace{-0.8cm}-{2Q_2\over7}(3\lambda^2+7\Lambda{\cal E}-5\Lambda
z\lambda)\bigg]\bigg\}\nonumber \ene

We now proceed to calculate the contribution to the centroid
frequency changes due to the induced adiabatic pressure change.
The adiabatic approximation is now well-justified on the grounds
that the layer where the field is expected is located deep enough.
Setting $\delta S=0$ in eq.\,[\ref{D_s_02}] and using
eq.\,[\ref{V_0_B_t}], to express ${\cal V}_0$, we obtain $$\Delta
D_{s,0}={1\over16\pi}\int dr r^2{\cal D}_{\rm ad}
(B_{t,1}^2+B_{t,2}^2).$$ Eq.\,[\ref{D_s_ad_0}] gives an explicit
form of ${\cal D}_{s,{\rm ad}}$ in terms of the eigenfunctions.

The corresponding contribution to the splittings is obtained from
inserting eqs.\,[\ref{V_1_B_t}] and [\ref{H_1_B_t}] into
eq.\,[\ref{Delta_D_sk}] yields

\bee D_{s,k}&=&{Q_k\over8\pi}\int
dr r^2 \bigg\{B_{t,1}^2\bigg[-\half({\cal D}_s^V+)\delta_{k1}\bigg]\nonumber\\
&&2{\cal D}_s^H+B_{t,2}^2\bigg[{1\over14}(5{\cal D}_s^V-10{\cal
D}_s^H)\delta_{k1}\nonumber\\&& -{3\over7}(2{\cal D}_s^V+3{\cal
D}_s^H)\delta_{k2}\bigg]\bigg\}. \nonumber \ene Combining all
three $D$ integrals into a single expression for the frequency
shift due the $j^{\rm th}$-component of the toroidal field, we
have
\begin{equation}
(\Delta\omega)_{t,j}={1\over8\pi I\omega}\sum_{k=0}^jQ_k\int
drr^2B_{t,j}^2{\cal R}_{t,kj}. \label{Delta_omega_tj}
\end{equation}
The $j^{\rm th}$ component generates all the $\gamma$'s from $k=0$ up to $j$.
From the $D$'s calculated above, we get for the following expressions
for the
 ${\cal R}$'s at $j=1$ and 2
$${\cal R}_{t,01}={\cal R}_{t,02}=\half{\cal D}_{\rm ad}+\lambda^2+{3\over2}\Lambda{\cal E},$$
$${\cal R}_{t,11}=-\half[{\cal D}_s^V+2{\cal D}_s^H+2\lambda^2+6\Lambda({\cal E}-z\lambda)],$$
$${\cal R}_{t,12}={5\over14}({\cal D}_s^V-2{\cal D}_s^H+2\lambda^2+6\Lambda z\lambda)$$
\bee
{\cal R}_{t,22}&=&-{1\over7}[6{\cal D}_s^V+9{\cal D}_s^H+12\lambda^2+
\nonumber\\&&4\Lambda(7{\cal E}-5z\lambda)]\nonumber\ene

The quantities ${\cal D}_{\rm ad}$, ${\cal D}_s^V$, and
${\cal D}_s^H$, are given in eqs.\,[\ref{D_s_ad_0}], [\ref{D_s_V}] and [\ref{D_s_H}],
respectively. In these equations, one may make use of yet another
approximation that is valid for p-modes in the acoustic propagation zone and
beneath. For solar p-modes (the f-modes are irrelevant here), we have
$\lambda\approx-\Lambda\eta z$, where
$$\eta\equiv\left({\omega\over{\cal L}_\ell}\right)^2
={1\over\Lambda}\left({\omega r\over c}\right)^2$$ is the square
of the ratio of the mode to Lamb frequencies. At the inner turning
point, we have $\eta=1$. With the above expression for $\lambda$,
we will obtain a more explicit form of the ${\cal R}$'s. The
products of the eigenfunctions occurring in the ${\cal R}$'s may
now be expressed as follows
\begin{equation}
z\lambda=-\eta{\cal
E}^H,\quad\lambda^2=\Lambda\eta^2{\cal E}^H,\quad
{\cal E}^H\equiv\Lambda z^2, \label{z_lambda}
\end{equation}
\begin{equation}
\zeta={\Lambda\omega^2r\over g}yz(2-\eta)=
{c^2\over gr}\Lambda^2\eta(2-\eta)yz. \label{Omega}
\end{equation}
Finally, we assume the ideal gas law, which is a fully adequate
approximation in the region considered, to obtain \bee {\cal
R}_{t,01}&=&{\cal R}_{t,02}\approx\Lambda\bigg[{11\over12}{\cal E}
-\bigg({\eta^2\over3}-2\eta\nonumber\\&&
+{3\over5}\bigg){\cal E}^H\bigg]\nonumber \ene
$${\cal R}_{t,11}\approx-\zeta-\Lambda\left[{\cal
E}+\eta\left(5-{2\eta\over3}\right){\cal E}^H\right]$$ $${\cal
R}_{t,12}\approx-{15\over7}\zeta+{5\over7}\Lambda\left[2{\cal
E}+\eta\left({8\over3}\eta-5\right){\cal E}^H\right]$$ $${\cal
R}_{t,22}\approx-{6\zeta\over7}-{\Lambda\over7}[10{\cal
E}-\eta(3\eta+38){\cal E}^H]$$ We now rewrite
eq.\,[\ref{Delta_omega_tj}] in the following convenient form,
specialized for the sun
\begin{equation}
\gamma_{t,k}=\sum_{j\ge k}\int_0^1 dxK_{t,kj}
\left({B_{t,j}\over\mbox{1 MG}}\right)^2\quad\mbox{$\mu$Hz}, \label{gamma_tk}
\end{equation}
with $x=r/R_\odot$
$$K_{t,kj}=1.48\times10^{-4}{\mbox{1 mHz}\over\nu}x^2{\cal R}_{t,k}.$$

\begin{figure}[]
\plotone{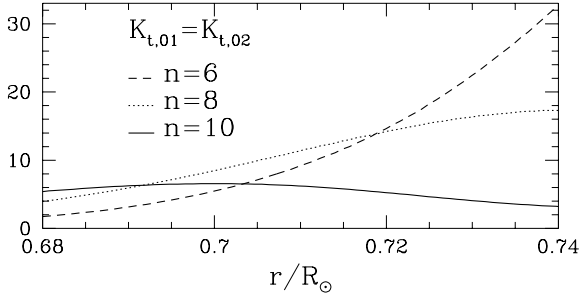}
 \caption{Kernels for
calculating the values of $\gamma_{t,0}$ according to
eq.\,[\ref{gamma_tk}], and arising from dipolar ($j=1$) and
quadrupolar ($j=2$) toroidal magnetic fields, for four $\ell=30$
modes of indicated orders $n$ in the zone near the bottom of the
convective envelope ($x=r/R_\odot=0.71$).}
\end{figure}

The modes that are most sensitive to the field in the vicinity of
the base of the convective envelope are those of moderate degree
with turning points located there. This is illustrated in Figs. 7
and 8, where we show the kernels $K_{t,k}(x)$ for three $\ell=30$
modes. The modes have frequencies in the 1.95 to 2.63 mHz range.
The lower turning points range, accordingly, from $x=0.742$ to
0.654. The $n$=6 mode, which has its inner turning point at
$x=0.721$ that is above the base of the convective zone, probes
the field not only within the convective envelope, but also the
region immediately beneath, which is below its inner turning
point!  This latter fact is in contrast to the ray approximation
in which this mode would know nothing about the region beneath its
inner turning point.  The turning of the $n=8$ mode is at $x=0.68$
and this mode is the best probe of the bulk of the overshooting
zone that extends down to about 0.70.  Similar results would apply
for different $\ell$-values, after an appropriate re-scaling  of
the frequencies so that the $\eta\propto\ell/\nu$ ratio is
preserved. We see that the toroidal field increase leads to a
corresponding increase in $\gamma_0$ and that the effect is mostly
of the opposite sign for $k>0$.

\begin{figure}[]
\plotone{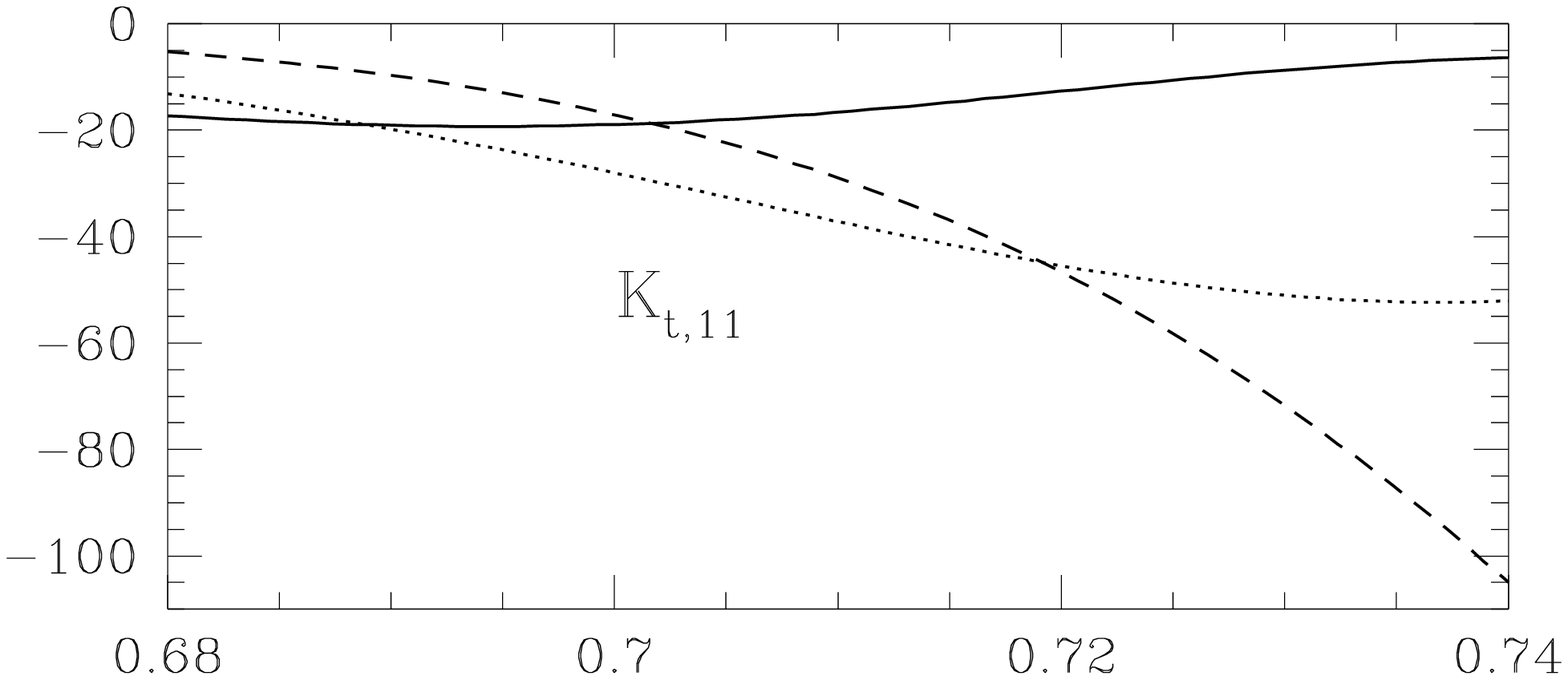} \plotone{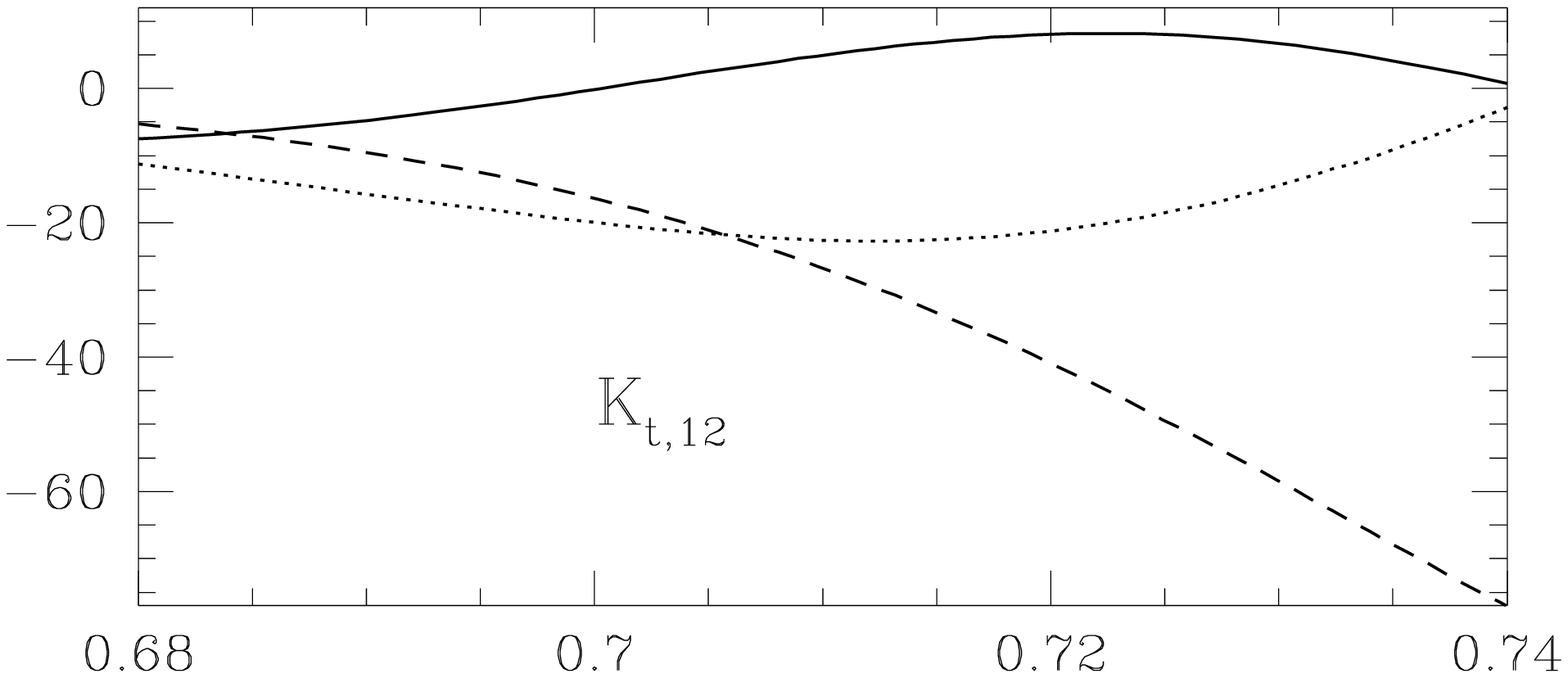} \plotone{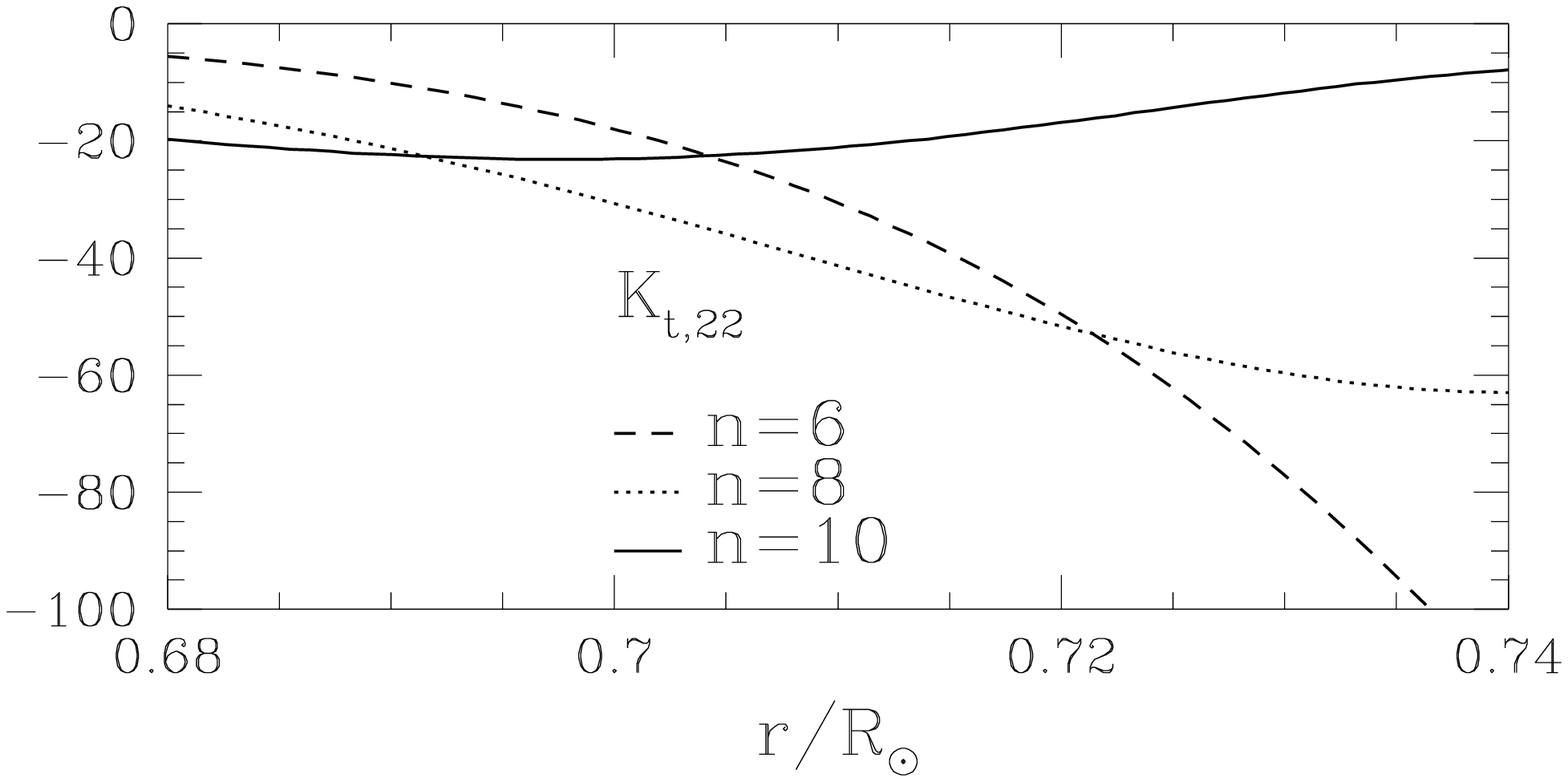}
 \caption{The same as Fig.7 but for the even-$a$ splittings ($k>0$).}
\end{figure}

If a  toroidal field of 1 MG would prevail in the layer between
$x=0.68$ and 0.74, that is over a distance comparable to one
pressure scale height, which is about 0.08, then the value of
$\gamma_0$ would reach up to $0.8 \mu$Hz, that of $\gamma_1$ would
be negative, reaching to -2.6, and that of $\gamma_2$ would also
be negative reaching $-2.9 \mu$Hz. Clearly, such values are very
significant and the field would be easily detectable. Detection of
the signal corresponding to a putative 0.1 MG field is problematic
at present day accuracy. The best chance is to see it is in the
even-$a$ coefficients, if indeed the field were dominated by the
low-$j$ polynomials. If the 1 MG field were present only within
the overshoot zone, extending from (say) x=0.65 to the base of the
convective zone, then the corresponding extreme values would be
-0.31, -0.95, and -1.1 $\mu$Hz.  Thus, somewhat stronger than 0.1
MG fields are required for detection. However, stronger fields may
be anticipated in the overshoot layer.

Chou and Serebryanskyi (2002) found evidence for a 0.4-0.7 MG
field at the base of the convective envelope from time-distance
seismology. Such a field, if it persists over a distance to
comparable to that assumed by us, should be detectable by means of
global seismology. However, the effort made so far did not result
in the detection of a significant signal (Basu, 2002).

\begin{figure}[ht]
\plotone{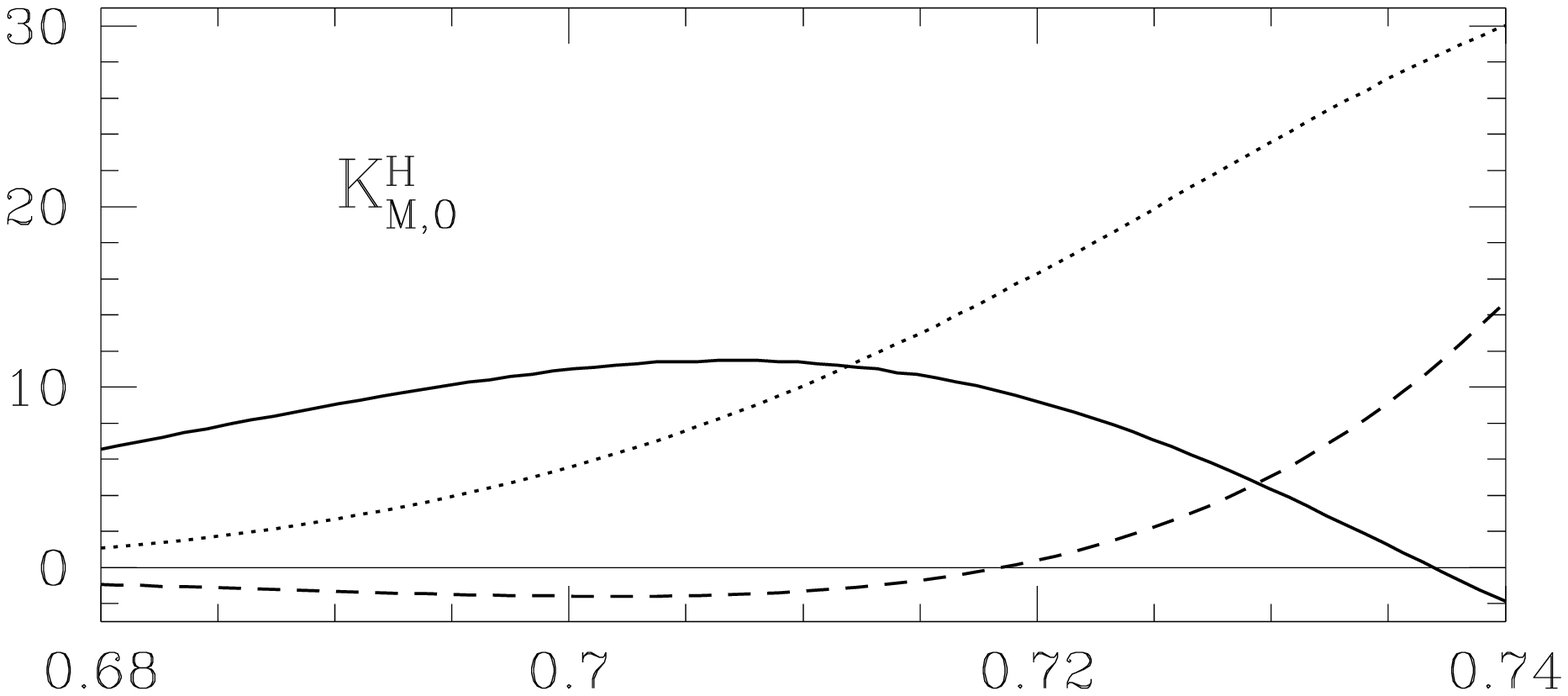} \plotone{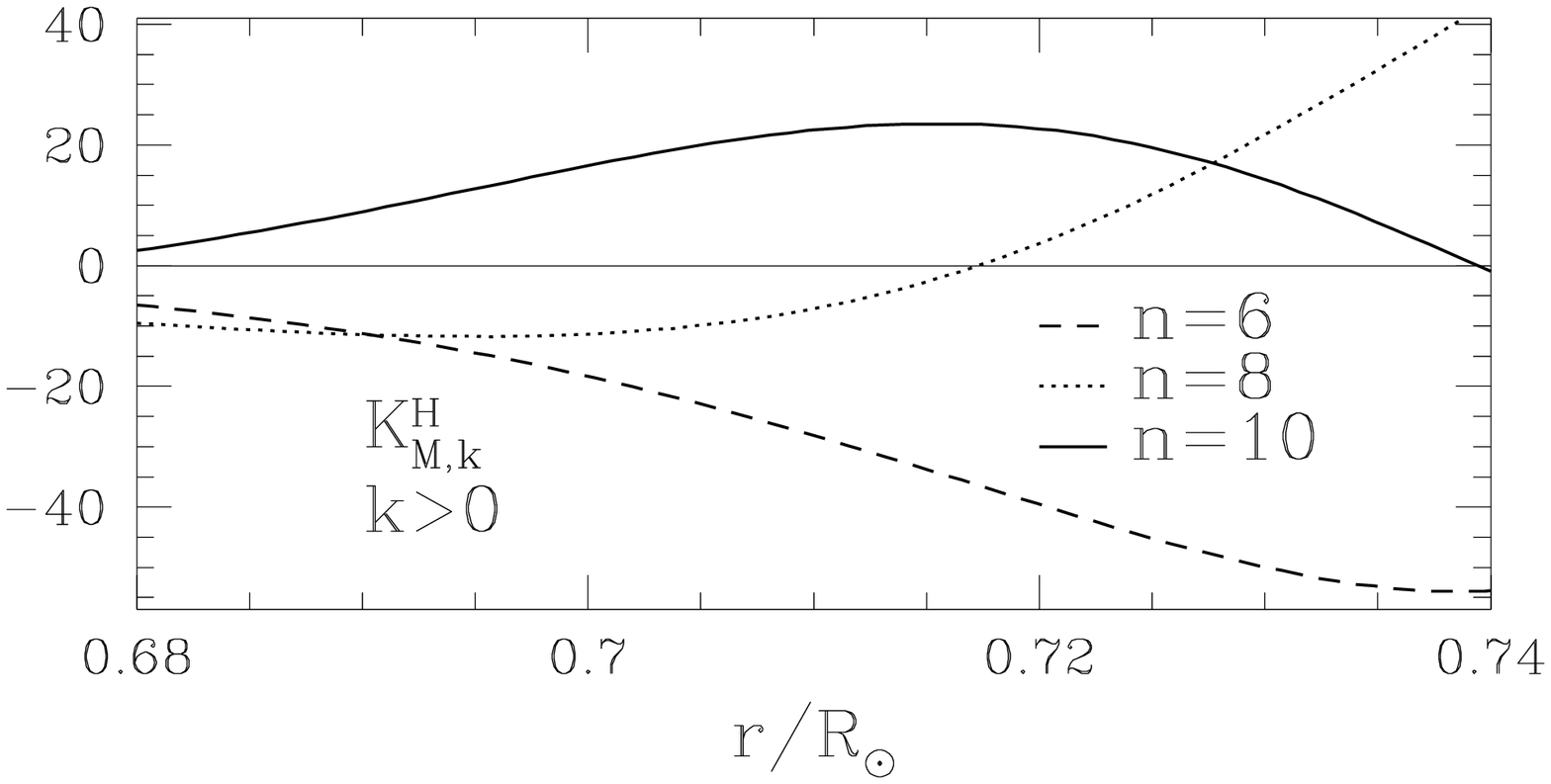} \caption{Kernels for
calculating the $\gamma$'s arising from small-scale horizontal
magnetic field near the base of the solar convective zone.}
\end{figure}
It is possible that the magnetic field in the deep part of the
convection envelope, and in the overshoot zone forms azimuthal
ropes, and thus is better represented as a small-scale field with
its mean value being a slowly varying function of latitude. In
this case, the frequency perturbation is described by the
adiabatic version of eqs.\,[\ref{Delta_omega_M_0}], at $k=0$, and
by [\ref{Delta_omega_Mk}], at $k>0$, with only the horizontal
components included. For the $\gamma$'s, we use an expression that
is similar to that given in eq.\,[\ref{gamma_M_k}]

\begin{equation}
\gamma_k^H=\int_0^1 dxK_k^H \left({B_{t,j}\over\mbox{1
MG}}\right)^2\quad\mbox{$\mu$Hz},
\end{equation}
with $x=r/R_\odot$
$$K_k^H=1.48\times10^{-4}{\mbox{1 mHz}\over\nu}x^2{\cal R}^H_{M,k}.$$
 With the approximation for the
eigenfunctions, which is valid for p-modes in the zone considered
as given in Eqs. [\ref{z_lambda}] and [\ref{Omega}], we have
$${\cal R}^H_{M,0}\approx\Lambda\left[{11\over5}{\cal
E}-\left({2\over3}\eta^2+ {6\over5}\right){\cal E}^H\right]$$
$${\cal R}^H_{M,k}\approx\Lambda\left[3{\cal E}+2\eta(\eta-2){\cal
E}^H\right]-2\zeta.$$ Both kernels change sign in the region of
interest.  In Fig. 9, we show examples of the kernels for the same
modes and in the same layer, as in Fig. 8. Differences between the
figures are apparent. Note in particular, the sign changes within
the layer. With a 1MG field in the layer, we get $\gamma_0$ of
about 0.8 $\mu$Hz for the $n=8$ and 10 modes. The absolute values
at $k>0$ are somewhat lower. So that a 0.4-0.7 MG field should be
easily detectable in the frequency shifts.  The expected signal in
the even-$a$'s would be somewhat weaker, but should also be
detectable.
\section{Conclusions}
We surveyed various effects that may explain the increase of the
mean frequencies of solar oscillations and the changes in the fine
structure of the multiplets correlated with the rise of activity.
Beyond any doubt is the fact that the seismic changes reflect
temporal and longitudinal evolution of the sun's activity and that
the main part of these changes has its source close to
photosphere, perhaps within 1 to 1.5 Mm. The question that remains
to be answered is how the changes are related to magnetic field
variations and associated variation of the velocity field and the
temperature in the atmosphere, and in the subphotospheric layers.
The answer is important because only after we know it, we will be
able to make a full use of seismic data as a probe of the
large-scale variability of the sun over its activity cycle.

Another open question is whether current frequency data reveal
changes in the deep interior. Assessing the intensity of the
magnetic field in the lower convective envelope and in the outer
radiative interior, just beneath, would be of great importance for
understanding the physics of magnetic activity.

With these questions in mind, we developed integral formulae
expressing shifts in oscillation frequencies in terms of changes
in magnetic and velocity fields and temperature. We made certain
approximations, which are well-justified both for solar p- and
f-modes. Small scale-fields were represented by the
square-averaged values of  vertical and horizontal components.
These values were expressed in terms of the Legendre-polynomial
series. We showed the coefficients at $P_{2k}$ in these series
contribute only to $a_{2k}$ coefficients of the frequency
splitting. The connecting expressions were given in the form of
integrals over the depth with explicit expressions for the
kernels. We also provided kernels linking the mean frequency
shifts and the $a_{2k}$ coefficients to a low-order expansion of a
putative large-scale toroidal field near the base of the
convective envelope.

Plots of the kernels allowed us to make a simple assessment of the
changes needed to explain the measured shifts. We found that the
increase of the mean frequencies and the changes in
$a_{2k}$-coefficients are most easily explained in terms of a
decrease of turbulent velocities associated with the increase of
the magnetic field with growing activity. The required decrease in
the turbulent velocity needed to explain the data constitutes only
a fraction of a percent. A decrease in turbulent velocity is
expected to result in a temperature decrease in outer layers of
the sun. Our estimate showed that the resulting temperature
decrease should give a significant contribution to the mean
frequency increase, which reduces requirement on the size of the
decrease of the turbulent velocity. Accounting for the seismic
changes by the sole direct effect magnetic field is more
difficult. An increase of the surface-averaged r.m.s. value of the
vertical field component by about 0.1 kG between the minimum and
maximum of the activity would be needed to account for the mean
frequency increase of the p-modes. The measured changes in the
even $a$ coefficient for p-mode require about twice as large field
increase at most active latitudes. Also a larger field seems to be
needed to account for the systematic increase of the f-mode
frequencies.

Considering the influence of the field near the base of the
convective envelope, we found that there is a chance for detecting
such a field directly from the frequency data. Evidence, from
time-distance seismology, for the presence there of the
$(0.4-0.7)$ MG field was recently put forward by Chou \&
Serebryanskyi (2002). We showed that such a field, if extends over
a layer of thickness comparable with one pressure distance scale,
should be detectable also by means of global seismology.

\acknowledgments

This research was supported in part by a Polish grant (KBN-5 P03D
012 20), U.S. grants from NASA (NAG5-12782) and NSF (ATM-0086999).


\section{Kernels for the $\gamma$'s}

Here we summarize the expressions for the kernels for evaluating
the $\gamma$'s due to small-scale velocity and magnetic fields. We
begin by recalling the definition of the {\it radial
eigenfunctions}:

$$ \vxi =r[y(r)\ve_r +z(r)\nab_H]\Y(\theta,\phi)
\exp(-{\rm i}\omega t).$$  We also define
$$\lambda= y{gr\over c^2}-z{\omega^2r^2\over c^2}$$
and
$${\cal E}=y^2+\Lambda z^2.$$

The kernels for {\it centroid shifts} ($\gamma_0$) due to
turbulent velocity are
$$
{\cal R}_{v,{\rm isoth}}^V= {\cal D}_{\rm
isoth}-(\lambda^2+2\Lambda z\lambda+\Lambda{\cal E}) \eqno(A1) $$

$${\cal R}_{v,0}^H=-{\Lambda\over2}{\cal E}.\eqno(A2)$$ and those
due to a small-scale magnetic fields are $${\cal R}_{M,{\rm
isoth}}^V=2\Lambda(z\lambda+2{\cal E})- {\cal D}_{\rm isoth}
\eqno(A3)$$

$${\cal R}_{M,{\rm isoth}}^H=2\lambda^2+\Lambda(4z\lambda+3{\cal
E})+ {\cal D}_{\rm isoth}.\eqno(A4)$$ where

\bee{\cal D}_{\rm isoth}&=&-\Gamma\bigg[\bigg(1+\Gamma_p+{1+\Gamma_\rho\over\chi_\rho}\bigg)\lambda^2
\nonumber\\&&+
{2\Lambda z\lambda\over\chi_\rho}\bigg]\nonumber
\ene

The kernels for the {\it even-$a$ coefficients} ($\gamma_k$, $k>0$)
due to turbulent velocity are
$${\cal R}_{v,k}^V=-(2\zeta-\psi+\lambda^2+2\Lambda z\lambda+\Lambda{\cal E}), \eqno(A5)$$
$${\cal R}_{v,k}^H=\zeta-\half[\psi-\Gamma(1+\Gamma_p)\lambda^2+2\Lambda z\lambda+3\Lambda{\cal E}].\eqno(A6)$$
and those due small-scale magnetic fields
$${\cal R}_{M,k}^V=4\zeta-2\psi-\Gamma(1+\Gamma_p)\lambda^2+2\Lambda{\cal E}, \eqno(A7)$$
$${\cal R}_{M,k}^H=-2\zeta+\psi+2\lambda^2, \eqno(A8)$$
where
$$\zeta={\omega^2r\over g}(y\lambda+2\Lambda zy)$$
and

\bee\psi&\equiv&\Gamma_\rho\bigg[\bigg(2-\Gamma{d\ln c^2\over d\ln
p}\bigg)\lambda^2+ 2\lambda\bigg(\Lambda z \nonumber\\&&
-{\omega^2r\over g}y\bigg)\bigg]-\Gamma{d\Gamma_\rho\over d\ln p}\lambda^2.
\nonumber\ene




\end{document}